\documentclass[3p,times,review]{elsarticle}

\usepackage[figuresright]{rotating}

\usepackage{stmaryrd,bm,amsmath}
\usepackage{dsfont}
\usepackage{amsbsy,amsmath,amsthm,latexsym,amssymb,mathtools}
\usepackage{graphicx,stmaryrd,sectsty}
\usepackage{setspace}
\usepackage[font=small,labelfont=bf]{caption}
\usepackage{epstopdf}
\usepackage{verbatim}
\usepackage{graphicx}
\usepackage{caption}
\usepackage{subcaption}
\usepackage{float}
\usepackage[toc,page]{appendix}
\usepackage{tikz}
\usepackage{enumitem}
\usepackage[utf8]{inputenc}

\newcommand{\bc}{\begin{center}}
\newcommand{\ec}{\end{center}}
\newcommand{\bfr}{\begin{flushright}}
\newcommand{\efr}{\end{flushright}}

\newcommand{\be}{\begin{enumerate}}
\newcommand{\ee}{\end{enumerate}}
\newcommand{\bi}{\begin{itemize}}
\newcommand{\ei}{\end{itemize}}
\newcommand{\bd}{\begin{description}}
\newcommand{\ed}{\end{description}}
\newcommand{\beq}{\begin{equation}}
\newcommand{\eeq}{\end{equation}}
\newcommand{\bea}{\begin{eqnarray}}
\newcommand{\eea}{\end{eqnarray}}

\newcommand{\bfi}{\begin{figure}}
\newcommand{\efi}{\end{figure}}
\newcommand{\bay}{\begin{array}{l}}
\newcommand{\eay}{\end{array}}

\newcommand{\mbf}{\mathbf}

\newcommand{\cref}[1]{(\ref{#1})}   

\begin{document}

\begin{frontmatter}

\title{Finite element modeling of dynamic frictional rupture with rate and state friction}

\author[l1]{Roozbeh Rezakhani \footnote{Corresponding author. \\ E-mail address: rrezakhani@u.northwestern.edu}}
\author[l1]{Fabian Barras}
\author[l3]{Michael Brun}
\author[l1]{Jean-Fran\c cois Molinari} 

\address[l1]{Civil Engineering Institute, Materials Science and Engineering Institute, Ecole Polytechnique Fédérale de Lausanne (EPFL), Lausanne,
Switzerland.}
\address[l3]{Univ Lyon, INSA-Lyon, GEOMAS, 69621 Villeurbanne, France.}

\begin{abstract}
Numerous laboratory experiments have demonstrated the dependence of the friction coefficient on the interfacial slip rate and the contact history, a behavior generically called rate and state friction. Although numerical models have been widely used for analyzing rate and state friction, in general they consider infinite elastic domains surrounding the sliding interface and rely on boundary integral formulations. Much less work has been dedicated to modeling finite size systems to account for interactions with boundaries. This paper investigates rate and state frictional interfaces in the context of finite size systems with the finite element method in explicit dynamics. It is shown that due to the highly non-linear nature of rate and state friction and its sensitivity to numerical noise, the time integration step to achieve an accurate steady state solution is orders of magnitude smaller compared to the stable time step required in boundary integral formulations. We provide evidence that the noise, which is source of instability in the finite element solution, originates from internal discretization nodes. We then investigate the long term behavior of the sliding interface for two different friction laws: a velocity weakening law, for which the friction monotonously decreases with increasing sliding velocity, and a velocity weakening-strengthening law, for which the friction coefficient first decreases but then increases above a critical velocity. We show that for both friction laws at finite times, that is before wave reflections from the boundaries come back to the sliding interface, a temporary steady state sliding is reached, with a well-defined stress drop at the interface. This stress drop gives rise to a stress concentration and leads to an analogy between friction and fracture. However, at longer times, that is after multiple wave reflections, the stress drop is essentially zero, resulting in losing the analogy with fracture mechanics. Finally, the simulations reveal that velocity weakening is unstable at long time scales, as it results in an acceleration of the sliding blocks. On the other hand, velocity weakening-strengthening reaches a steady state sliding configuration.
\end{abstract}


\end{frontmatter}
\section{Introduction}	
Friction is an important ingredient controling the nucleation and propagation of dynamic shear ruptures at contact interfaces. Various forms of frictional laws have been formulated among which are the slip-weakening and the rate and state friction laws. In slip-weakening friction, the coefficient of friction at any point along the interface depends on the relative slip at that point, and decreases from a static value to a dynamic one as the relative slip increases \cite{andrews_rupture_1976}. The subtle evolution of the friction coefficient measured experimentally during frictional sliding is better described by rate and state laws, for which the coefficient of friction depends on the slip velocity as well as a state variable, which is intended to describe the contact duration (state) of the micro contacts. It has been shown with fault laboratory experiments that rupture nucleation and propagation can be captured with rate and state friction, but not with the too simplistic slip-weakening friction \citep{rubino_understanding_2017}. Marone \cite{marone_laboratory-derived_1998} presented a complete review of rate and state friction laws in relation to interpreting laboratory experimental data on rock friction. Bizzarri and Cocco \cite{bizzarri_slip-weakening_2003} studied the evolution of slip velocity and shear traction distribution around the tip of a propagating shear rupture on interfaces controlled by rate and state friction. At a larger scale, rate and state friction has been extensively used to analyze and explain different aspects of earthquake scenarios \cite{scholz_earthquakes_1998,dieterich_constitutive_1994,lapusta_nucleation_2003}. Instability of rate and state friction and the corresponding critical nucleation length have been the topic of several recent works \cite{brener_unstable_2018,brener_dynamic_2016,aldam_critical_2017,viesca_self-similar_2016}.

Different features of dynamic rupture propagation on interfaces controlled by rate and state friction have been studied. It has been shown both in experiments and simulations that the dynamic rupture propagation mode can be either crack-like or pulse-like \cite{lu_pulse-like_2010, lu_rupture_2010, gabriel_transition_2012, zheng-rice-1998}, with characteristic propagation speeds \cite{anderson_fracture_mechanics, andrews_rupture_1976}. In crack-like ruptures, frictional sliding continues in the wake of the front, while the two surfaces immediately re-stick behind pulse-like ruptures. Several relevant works have investigated the influence of the friction law parameters, applied boundary conditions, nucleation procedures, and interface heterogeneities on the rupture mode and the propagation speed of the frictional rupture front \cite{coker_frictional_2005,shi_properties_2008,lu_analysis_2009,kaneko_spectral_2008,ampuero_cracks_2008}. Dunham et al. \cite{dunham_earthquake_2011} and Erickson et al. \cite{erickson_finite_2017} investigated the effect of off fault plasticity on gelogic faults governed by rate and state friction as well as the effect of bulk viscoelasticity \cite{allison_earthquake_2018}. Recent theoretical and numerical works uncovered under which conditions frictional sliding fronts can be described as classical rupture fronts, i.e. using a fracture-like energy balance. \cite{Barras_partI,Barras_partII}. An essential condition is the existence of a well-defined drop of frictional stress behind the rupture front, for which the rate-dependent aspects of friction are crucial.. 

Numerical modeling of frictional sliding has been under increased scrutiny. The Boundary Integral Method (BIM) \cite{geubelle_spectral_1995} has been widely used to simulate rupture propagation on frictional interfaces \cite{barras_supershear_2018,barras_interplay_2017,lu_pulse-like_2010,lu_rupture_2010}. Although computationally efficient, BIM is mostly limited to planar interfaces in infinite domains. Modeling various important aspects affecting interfacial friction among which are bulk inelasticity and heterogeneities, finite size effects, and interfacial damage zones is challenging if not out of reach with BIM. Considering the importance of the above mentioned factors for a realistic analysis of laboratory data and development of robust earthquake models, other numerical methods should also be explored. The Finite Element Method (FEM) \cite{belytschko_nonlinear_2014} can incorporate material nonlinearity, non planar faults, and finite-size effects including realistic domain boundaries. FEM has been extensively used for the simulation of frictional interfaces governed by Coulomb and slip-weakening laws \cite{kammer_length_2016,kammer_existence_2014}. Baillet et al. \cite{baillet_finite_2005} investigated dynamic instabilities in frictional contact interfaces between elastic and rigid solids and showed that steady state slip pulses can propagate even in the presence of a constant coefficient of friction. Recently, Tonazzi et al. \cite{baillet_instability_2013} studied different modes of dynamic instabilities between two elastic bodies. They presented how structural damping influences the global instability mode varying from stick-slip phenomenon to stable slip. The finite element method is also used for
the simulation of engineering applications involving frictional behavior such as braking processes \cite{baillet_instability_2007}. However, limited studies can be found on finite element modeling of rate and state dependent frictional interfaces. Laursen et al. developed a finite element framework to simulate rate and state friction using an implicit time integration scheme \cite{laursen_constitutive_1997,oancea_dynamics_1996}. Recently Tal and Hager \cite{tal_dynamic_2018} published theoretical formulations and computational steps of a mortar-based implicit finite element method to simulate frictional interfaces with special focus on rate and state frictional laws. To avoid convergence issues, which is characteristic of implicit schemes, an explicit time integration formulation can be adopted, which also simplifies the numerical procedure. Coker et al. \cite{coker_frictional_2005} and later Shi et al. \cite{shi_properties_2008} used an explicit dynamic finite element method to simulate rate and state frictional interfaces and to investigate the mode and speed of the propagating rupture under different loading conditions. 

This paper aims to clarify the physical concepts and constraints related to rate and state friction. We will focus on the popular pure velocity weakening (VW) friction and on the more physically realistic velocity weakening-strengthening (VWS) friction, which are explained in Section \ref{R&S-formulation}. Within the explicit dynamic finite element framework, Section \ref{FE-fric} considers  a frictional interface between two homogeneous elastic bodies. Details of time discretization procedure, calculation of normal contact and tangential frictional tractions in accordance with rate and state friction, and the numerical steps are presented in Appendices. Steady state sliding of the frictional interface governed by rate and state friction along with its numerical stability and time step requirement are presented in Section \ref{steady-state}. Section \ref{rupture-propagation} compares the FE results with BIM simulations for a problem in which a perturbation of the state variable leads to frictional rupture nucleation and propagation on the interface. It is shown that the interfacial slip velocity evolution obtained from FEM matches the BIM results, but at a higher computational cost. Finally, the analysis is reproduced for a finite-size domain, to highlight the key role of reflected waves at system boundaries to obtain a long term steady state sliding. A global steady state equilibrium is reached for VWS friction but not for VW friction, which is inherently unstable. 

\section{Rate and state dependent friction law}	\label{R&S-formulation}
In rate and state friction, the friction coefficient $f(v(x,t),\phi(x,t))$ depends on the relative slip velocity $v(x,t)$ and a state variable $\phi(x,t)$. The state variable has dimension of time and intends to describe the average lifetime of load-carrying microscopic contact asperities. It evolves with static contact time duration, the relative slip velocity between the sliding solids, and the normal stress on the interface. Rate and state friction formulation, originally derived by Dieterich \cite{dieterich_modeling_1979} and shortly followed by Ruina \cite{ruina_slip_1983}, expresses the friction coefficient as a logarithmic function of slip velocity and state variable 
\begin{equation}\label{RS-VW}
f(v,\phi) = f_0 + a \ln\left(\frac{v}{v^*} \right) + b \ln\left(\frac{\phi}{\phi^*} \right)
\end{equation}
\begin{equation}\label{state-diff-eq}
\dot{\phi} = 1 - \frac{v \phi}{D}
\end{equation}
in which $f_0$, $a$ and $b$ are constants, which can be determined using experimental data. $v^*$ and $\phi^*$ are normalizing constants. Note that space $x$ and time $t$ dependence of slip velocity and state variable in Equations \ref{RS-VW} and \ref{state-diff-eq} are dropped in the rest of the paper to condense notation. The ordinary differential equation presented in Equation \ref{state-diff-eq} governs the time evolution of the state variable. If the interface is in sticking condition, the slip velocity is equal to zero, $v=0$, and Equation \ref{state-diff-eq} leads to $\phi = t$. This implies that the state variable increases with the stationary contact time, and in combination with Equation \ref{RS-VW}, the friction coefficient increases with logarithm of time consistent with experimental observations \cite{Dieterich_direct_observ_1994}. In steady state sliding, the state variable is constant, which implies 
\begin{equation}\label{phi_ss}
\dot{\phi} = 0 \Longrightarrow \phi_{ss} = D/v_{ss}
\end{equation}
in which $\phi_{ss}$ and $v_{ss}$ are respectively the state variable and slip velocity in the steady state sliding condition. Equation \ref{phi_ss} states that $\phi_{ss}$ is inversely proportional to the sliding velocity, which is rationalized by the decrease of the microcontacts lifetime with increasing slip velocity. At steady state sliding, the friction coefficient $f_{ss}$ corresponding to $v_{ss}$ is
\begin{equation}\label{fvw_ss}
f_{ss} = f_0 + a \ln\left(\frac{v_{ss}}{v^*} \right) + b \ln\left(\frac{D}{v_{ss}\phi^*} \right)
\end{equation}

\begin{figure}[t!]
	\centering 
	\includegraphics[width=0.8\textwidth]{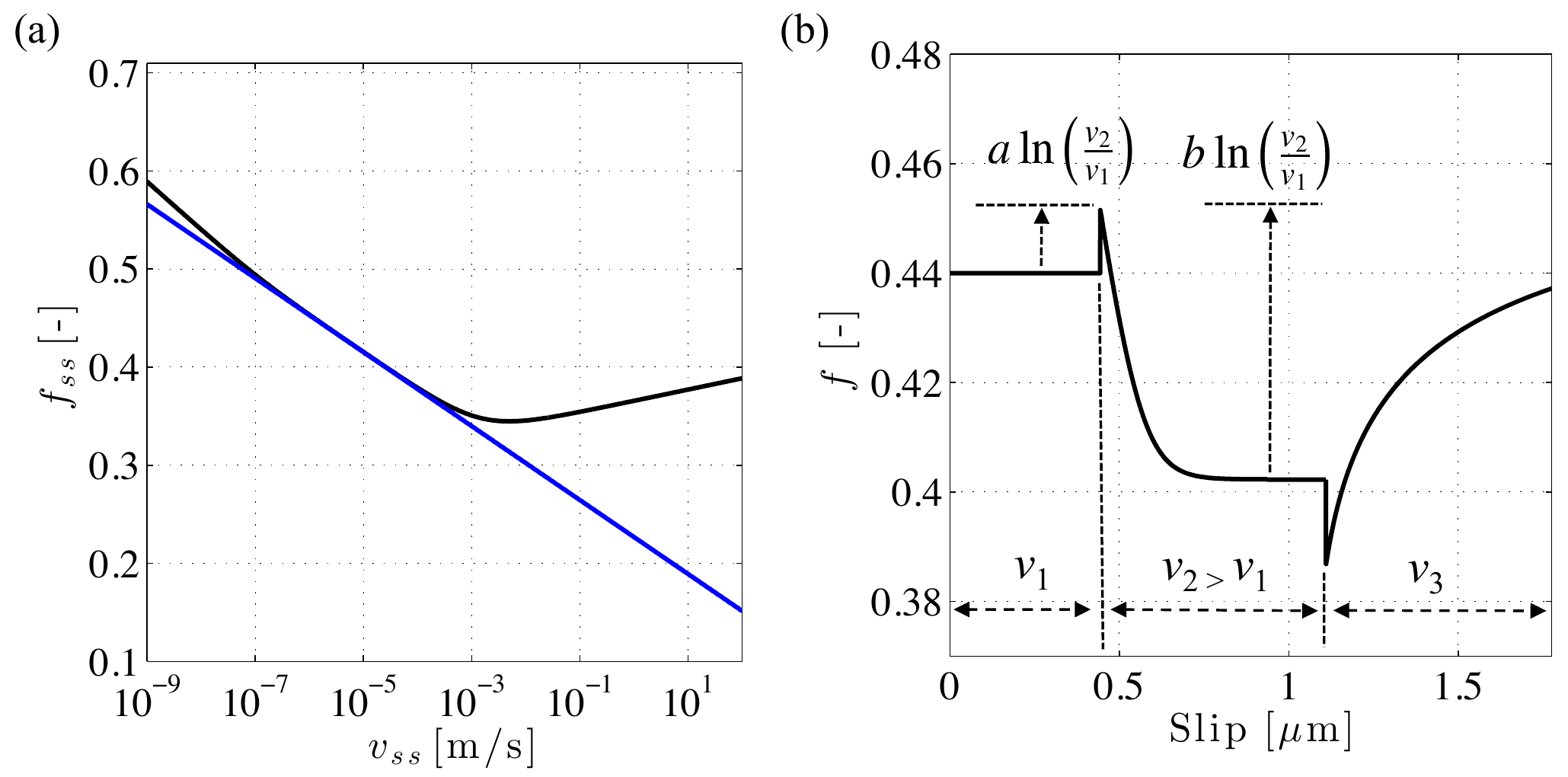}
	\caption{(a) Steady state friction coefficient versus steady state sliding velocity. The blue and black curves are the steady state form of VW and VWS presented in Equations \ref{RS-VW} and \ref{RS-VWS}, respectively. (b) Evolution of friction coefficient due to a sudden change in sliding velocity for VW friction. }
	\label{fricjump-steadystate}
\end{figure}

$f_{ss}$ versus $v_{ss}$ is plotted in Figure \ref{fricjump-steadystate}a in blue curve, according to Equation \ref{fvw_ss} and using: $f_0 = 0.285$, $a = 0.005$, $b = 0.0214$, $v^* = 10^{-7}$ [m/s]; $\phi^* = 3.3 \times 10^{-4}$ [s]; $D = 5 \times10^{-7}$ [m]. For this set of parameters ($a<b$), the steady state coefficient of friction decreases as the steady state sliding velocity increases, which is called velocity weakening. This behavior can be converted to velocity strengthening by choosing ($a>b$). In fact, the change of $f_{ss}$ between two steady state sliding conditions $v_1$ and $v_2$ denoted as $\Delta f_{ss} = f^1_{ss} - f^2_{ss} = (a-b)\ln(v_1/v_2)$ is a positive value for ($v_2>v_1$) and $(a-b)<0$. Between two steady state sliding conditions, the interplay between Equations \ref{RS-VW} and \ref{state-diff-eq} governs the variation of friction coefficient. In Figure \ref{fricjump-steadystate}b, the variation between $v_1 = 2.22$ [$\mu$m/s] and $v_2 = 22.2$ [$\mu$m/s] ($v_2>v_1$), and between $v_2$ and $v_3 = 1$ [$\mu$m/s] ($v_3<v_1<v_2$) is depicted. One can see that a sudden jump in $f$ occurs as the sliding velocity changes, and it gradually evolves to the new steady state sliding value.  

A revised version of the rate and state fricition law was proposed by Dieterich \cite{Dieterich_1986,Okubo_Dieterich_1986} to describe the constitutive behavior of geologic faults. Shibazaki and Iio \cite{shibazaki_physical_2003} and recently Sinai et al. \cite{sinai_velocity-strengthening_2014} investigated this law in details, which is
\begin{equation}\label{RS-VWS}
f(v,\phi) = f_0 + a \ln\left( 1+\frac{v}{v^*} \right) + b \ln\left( 1+\frac{\phi}{\phi^*} \right)
\end{equation}
whose steady state form can be written as 
\begin{equation}\label{fvws_ss}
f_{ss} = f_0 + a \ln\left(1+\frac{v_{ss}}{v^*} \right) + b \ln\left(1+\frac{D}{v_{ss}\phi^*} \right)
\end{equation}

The steady state curve of this friction model is plotted in Figure \ref{fricjump-steadystate}a in black using the same parameters presented above. The friction coefficient displays a velocity strengthening response following the initial velocity weakening branch. While other versions of rate and state friction are available in the literature \cite{shi_properties_2008,nagata_revised_2012}, we restrict our analysis to the two friction laws presented above, and we compare them in the numerical simulations presented in Section \ref{rupture-propagation}. These reveal that, for both laws, a temporary steady state sliding is reached on the interface behind the propagating rupture front. However, their behaviors differ in the long term, that is after the reflected waves from the boundaries travel back to the sliding interface. We show that the velocity weakening law is unstable, and results in acceleration of the sliding blocks, while the velocity weakening-strengthening law reaches a global steady state sliding configuration of the blocks, and appears more physical for numerical modeling of friction. 

\section{Numerical modeling of the frictional interface}\label{FE-fric}
\begin{figure}[t!]
	\centering 
	\includegraphics[width=0.9\textwidth]{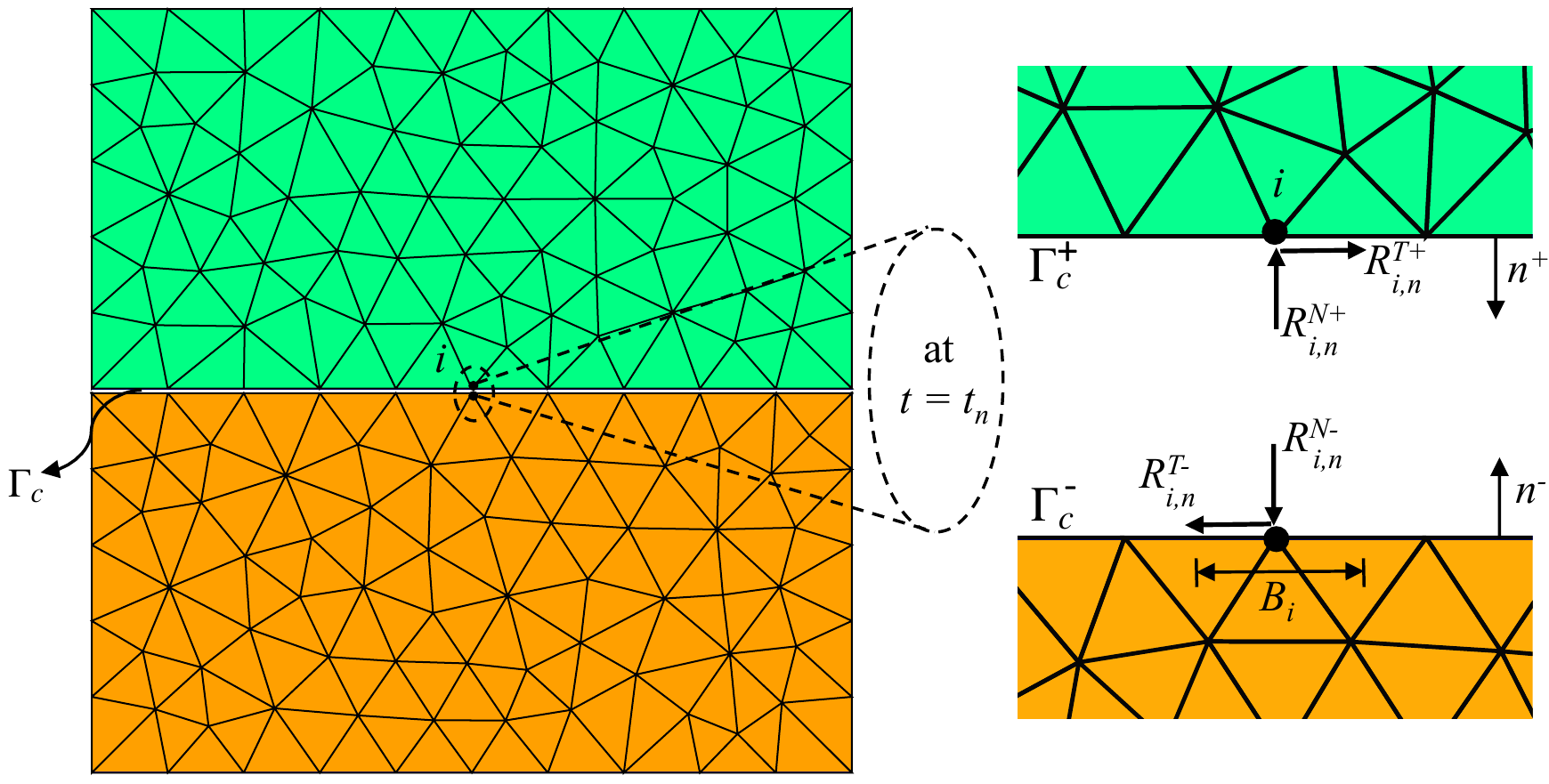}
	\caption{Finite element representation of two solid blocks that are in contact along $\Gamma_c$. The two blocks are discretized with conforming mesh on the interface, and each two corresponding nodes are grouped to apply contact and frictional constraints. A zoomed view of the forces acting among the pair of nodes $i$ is illustrated on the right at time $t = t_n$.}
	\label{FE-contact-setup}
\end{figure}

Frictional contact between two elastic bodies is simulated within the explicit dynamic finite element framework. Consider two elastic solids brought into contact as shown in Figure \ref{FE-contact-setup}. Conservation of momentum, is expressed in partial differential form as follows
\begin{equation} \label{motion-1}
\sigma_{ij,j} + b_i = \rho \ddot{u}_i
\end{equation}
where $\sigma_{ij}$ is the symmetric Cauchy stress tensor; $b_i$ is the body force vector per unit volume that will be dropped for simplicity; $\rho$ is the mass density; $u_i = u_i(x_k,t)$ is the time dependent displacement vector of any point in the material domain with position vector $x_k$. As the numerical examples presented in this paper are in 2D, $i,j,k = 1,2$. Boundary conditions for Equation \ref{motion-1} are in the form of displacement $u_i = U_{i}$ on $\Gamma_u$ and traction $\sigma_{ij}n_j = T_{i}$ on $\Gamma_t$. $\Gamma_u$ and $\Gamma_t$ are the portions of the domain boundary on which displacement $U_{i}$ and traction $T_{i}$ boundary conditions are applied, respectively. In addition, the initial displacement and velocity conditions are: $u_i(x_k,0) = U_{0i}(x_k)$ and $\dot{u}_i(x_k,0) = V_{0i}(x_k)$ defined on the whole material domain $\Omega$. Equation \ref{motion-1} can be converted to its weak form through the principle of virtual work, which states 
\begin{equation} \label{motion-2}
\int_{\Omega} \sigma_{ij} \delta u_{i,j} ~\text{d}\Omega + \int_{\Omega} \rho \ddot{u}_i \delta u_{i}~\text{d}\Omega =  \int_{\Gamma_t} T_i \delta u_{i} ~\text{d}{\Gamma_t} + \int_{\Gamma_c} R_i \delta u_{i} ~\text{d}{\Gamma_c}
\end{equation}
in which $\Gamma_c$ is the interfacial boundary as shown in Figure \ref{FE-contact-setup}, and $R_i$ is the vector of contact and frictional tractions on $\Gamma_c$. 

To solve Equation \ref{motion-2} numerically, the material domain is discretized by finite elements. The displacement vector at any point in each finite element is formulated in terms of the element nodal displacements and shape functions. We will employ either linear interpolation triangular elements (T3) or bilinear quadrilateral elements (Q4). Using this relation in Equation \ref{motion-2} along with some algebraic matrix manipulation, one obtains 
\begin{equation} \label{motion-3}
\mbf{M} \ddot{\mbf u} + \mbf{f}^{\text{int}} = \mbf{f}^{\text{ext}} +  \mbf{f}^{\text{cont}} + \mbf{f}^{\text{fric}}
\end{equation}
where $\mbf{u}$ is the displacement vector including all components of the nodal displacements. $\mbf{M}$ is the diagonal mass matrix of the discretized domain, which leads to a considerable reduction of the computational cost. $\mbf{f}^{\text{int}}$ and $\mbf{f}^{\text{ext}}$ are, respectively, the internal and external force vectors, refer to Reference \cite{belytschko_nonlinear_2014} for more details. $\mbf{f}^{\text{cont}}$ is the contact force vector normal to the interface, and $\mbf{f}^{\text{fric}}$ is the friction force vector tangential to the interface. The central difference method is employed to solve Equation \ref{motion-3}, with implementation details discussed in Appendix \ref{time-disc-proc}. In addition, derivation of $\mbf{f}^{\text{cont}}$  and $\mbf{f}^{\text{fric}}$ are presented in Appendices \ref{contact-force-derivation} and \ref{friction-force-derivation}. Finally, numerical calculations that are performed during each time step are reported in Appendix \ref{num-steps}. 

\section{Steady state sliding of the frictional interface}\label{steady-state}
\begin{figure}[t!]
	\centering 
	\includegraphics[width=\textwidth]{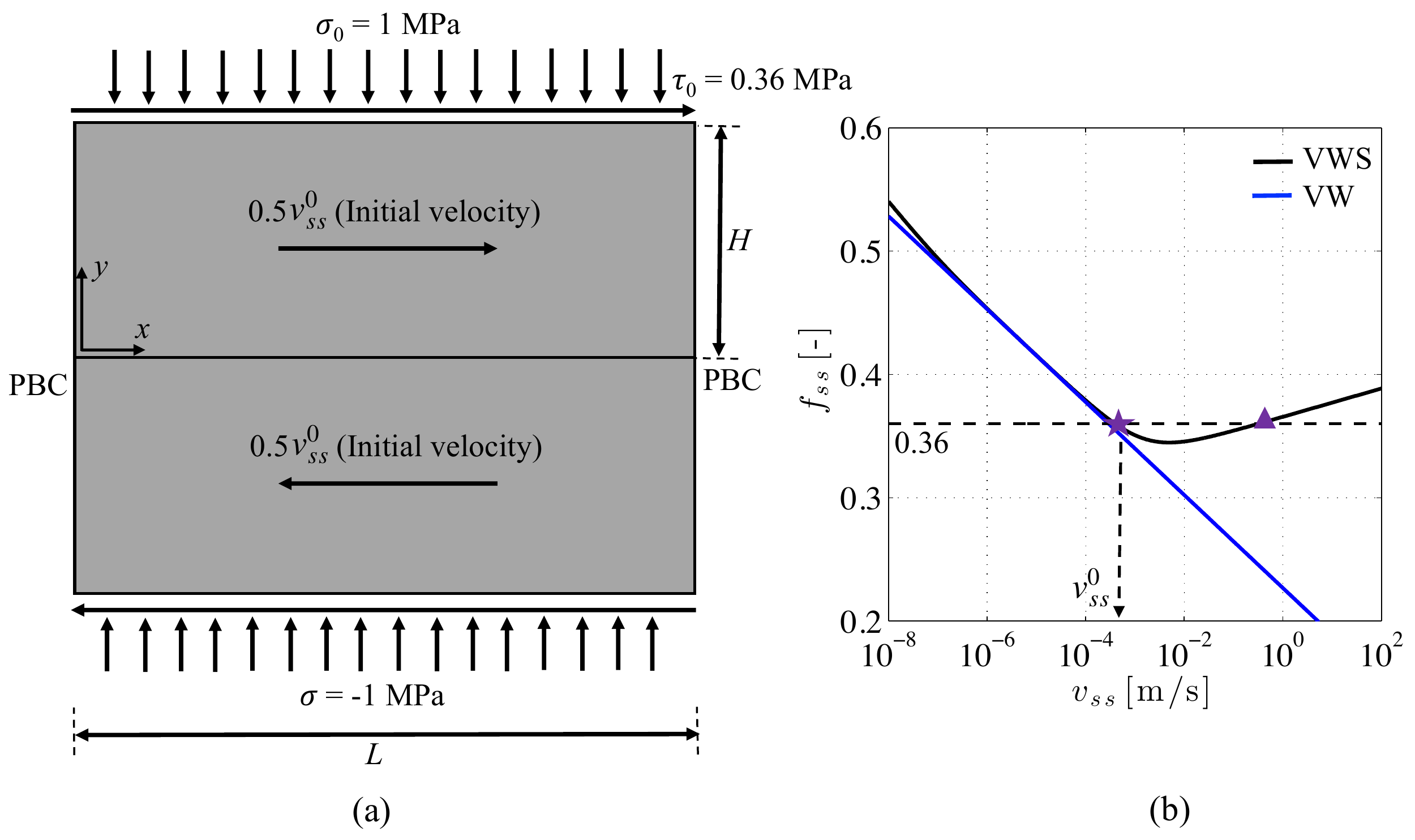}
	\caption{(a) Two elastic solids in contact and under normal and tangential far field tractions. Periodic boundary conditions are applied on lateral sides, and $0.5v_{ss}^0$ is the initial velocity assigned to both solids in opposite directions. (b) Steady state friction coefficient as function of steady state sliding velocity for the velocity weakening-strengthening friction (VWS), black curve, and the pure velocity weakening (VW) friction, blue curve. The purple star shows the initial steady state slip velocity at the interface and the corresponding friction coefficient. The purple triangle shows another point on the VWS curve, at which the friction coefficient value equilibrate the far field shear stress.}
	\label{elas-elas-setup-curve}
\end{figure}

\subsection{Set-up of the steady state sliding problem}\label{steady-state-setup}
We begin by investigating the steady state sliding, with an initial interfacial slip velocity of $v_{ss}^0$, of two solid blocks simulated by finite elements. Consider two elastic blocks of height $H=0.125$ and length $L=0.5$ [m] that are in contact as depicted in Figure \ref{elas-elas-setup-curve}a. The blocks are under normal pressure $\sigma_0$ and tangential stress $\tau_0$ on the upper and lower boundaries, while periodic boundary conditions are applied on the lateral sides. To reach steady state slip velocity of $v_{ss}$ over the interface, the initial velocities of the two solid blocks, $v(x,y,0)$, must be set to $0.5 v_{ss}^0$ and in opposite directions.

To obtain $v_{ss}^0$ corresponding to the applied $\tau_0$, the variation of the steady state friction coefficient in terms of the steady state sliding velocity should be considered, see Figure \ref{elas-elas-setup-curve}b. We begin by considering the velocity weakening-strengthening friction law shown as a black curve in Figure \ref{elas-elas-setup-curve}b. The pure velocity weakening law, the blue curve, will be considered in Section \ref{perturbation-VW}. In the current example, we prescribe the far field applied shear stress $\tau_0 = 0.36$ [MPa] and the normal stress $\sigma_0= -1$ [MPa]. Therefore, to preserve the initial applied velocity and maintain steady state sliding, force equilibrium in $x$ direction must be enforced, implying that $f_{ss} = \tau_0/\sigma_0 = 0.36$ should be achieved everywhere at the interface. In Figure \ref{elas-elas-setup-curve}b, the horizontal line depicts $f_{ss} = 0.36$, which intersects the steady state friction curve at two distinct points marked as purple star and triangle, corresponding to the sliding velocities of $3.8931 \times 10^{-4}$ and $0.32037$ [m/s] respectively . Any of these two slip velocities can be used as the initial $v_{ss}^0$ and will lead to a steady state sliding by satisfying the equilibrium condition. However, as we will discuss later the first intersection point (star) on the weakening branch is inherently unstable, as an increase of the slip velocity results in a decrease of friction and a further acceleration. Bulk material properties of the solids are $\rho = 1200$ [kg/m$^3$], $E = 0.8$ [GPa], and $\nu = 0.33$ resembling mechanical characteristics of PMMA. It is assumed that the problem is in plane stress condition. 

\subsection{Analysis of time step requirement}\label{time-step-study}
\begin{figure}[t!]
	\centering 
	\includegraphics[width=\textwidth]{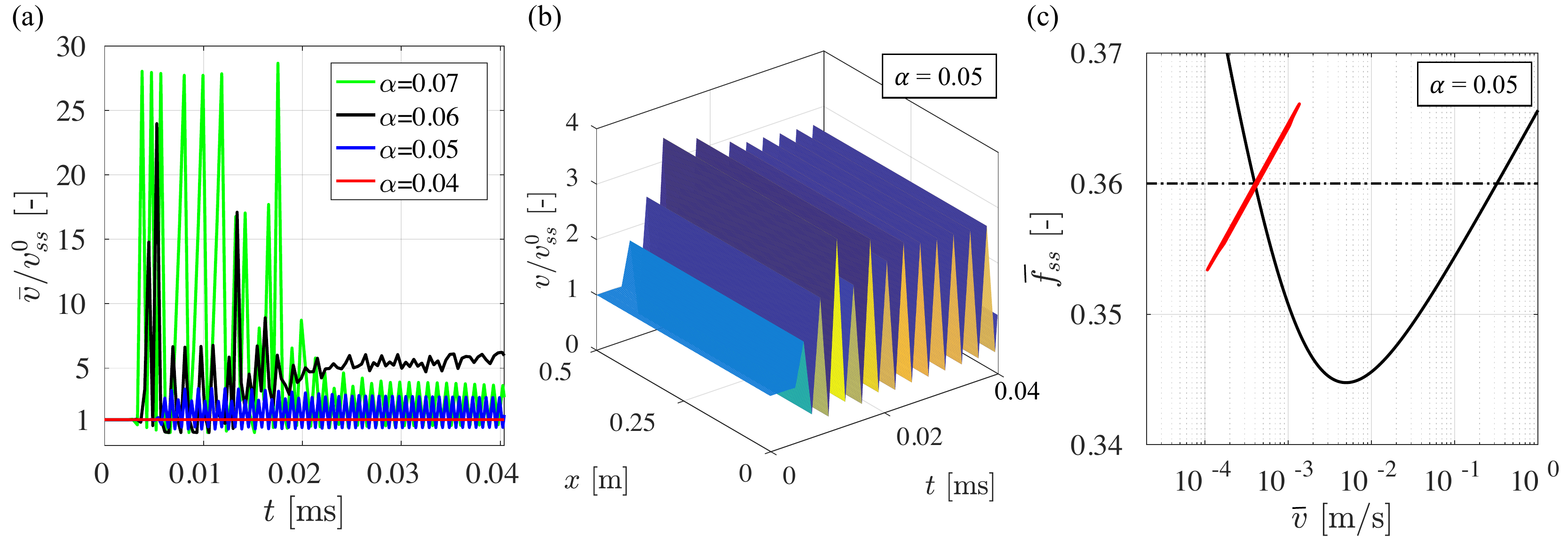}
	\caption{(a) Average velocity velocity over the interface normalized by the initial steady state velocity versus time for different $\alpha$ values. (b) Slip velocity over the interface plotted in time for $\alpha=0.05$, note that not all time steps are shown for clarity. (c) Variation (red color) of the interface average coefficient of friction with respect to the steady state curve for $\alpha=0.05$.}
	\label{250-instability}
\end{figure}

One of the main challenges in explicit finite element modeling of rate and state friction is the requirement of a very small time step compared to the stable time step calculated by the Courant–Friedrichs–Lewy CFL condition. Although this is a crucial issue, which considerably affects numerical accuracy of results including the physics of frictional rupture, as well as the computational cost, it has not been investigated in details in literature.
 Here, we analyse the effect of the chosen time step in achieving steady state sliding. We start by discretizing the interface with 250 elements, and the same element size is used to mesh the bulk with uniform linear quadrilateral elements. The critical time step according to CFL condition for the given mechanical properties and the finite element size is ${\Delta t}_{\text{CFL}} = 2.256$ [$\mu$s]. The problem is solved for different time steps calculated as ${\Delta t} = \alpha \times {\Delta t}_{\text{CFL}}$. In generic explicit finite element analysis, $\alpha$ is set to 0.5-0.8, while in the current study a much smaller number is necessary to achieve the correct response. The steady state simulation is performed for the simulation time of 0.04 [ms] with different values of $\alpha = 0.04,0.05,0.06,0.07$, and the evolution of slip velocity on the interface with the initial steady state slip velocity $v^0_{ss} = 3.8931 \times 10^{-4}$ [m/s] is recorded. It is expected that, in the absence of any sort of perturbation, the initial steady state slip velocity is preserved, and the whole interface slides with $v^0_{ss}$ till the end of the simulation. To study the effect of $\alpha$, the slip velocity is averaged over the entire interface at every time step and normalized with respect to the initial slip velocity, $\bar{v}/v^0_{ss}$, which is plotted versus time in Figure \ref{250-instability}a. One can see that for  $\alpha = 0.04$, a value of $\bar{v}/v^0_{ss}=1$ is retained, which means that the steady state sliding with $v^0_{ss}$ is accomplished throughout the simulation. On the other hand, for $\alpha =0.05,0.06,0.07$, an instability takes place after a certain time of steady state sliding, and $\bar{v}/v^0_{ss}$ starts to oscillate. One can also notice that the maximum amount of the oscillation is higher for higher values of $\alpha$. To understand how the velocity profile evolves on the interface and the instability mode, the interface slip velocity normalized by the initial steady state value ${v}/v^0_{ss}$ is plotted in space and time for $\alpha=0.05$ in Figure \ref{250-instability}b. One can see that ${v}/v^0_{ss}$ is equal to 1 in the initial steady state sliding phase, after which it starts to oscillate to other values uniformly over the whole interface. Note that in this plot, ${v}/v^0_{ss}$ is not plotted for all time steps for the sake of clarity, so not all oscillations are seen. These oscillations originate from an non-physical instability, as seen in the variation of the friction coefficient with slip velocity presented in Figure \ref{250-instability}c. The average interface coefficient of friction oscillates around the initial steady state location on the $f_{ss}-v_{ss}$ curve. Therefore, the choice of time step used in the explicit algorithm is of paramount importance and should be carefully examined for each specific example. 

\begin{figure}[h!]
	\centering 
	\includegraphics[width=\textwidth]{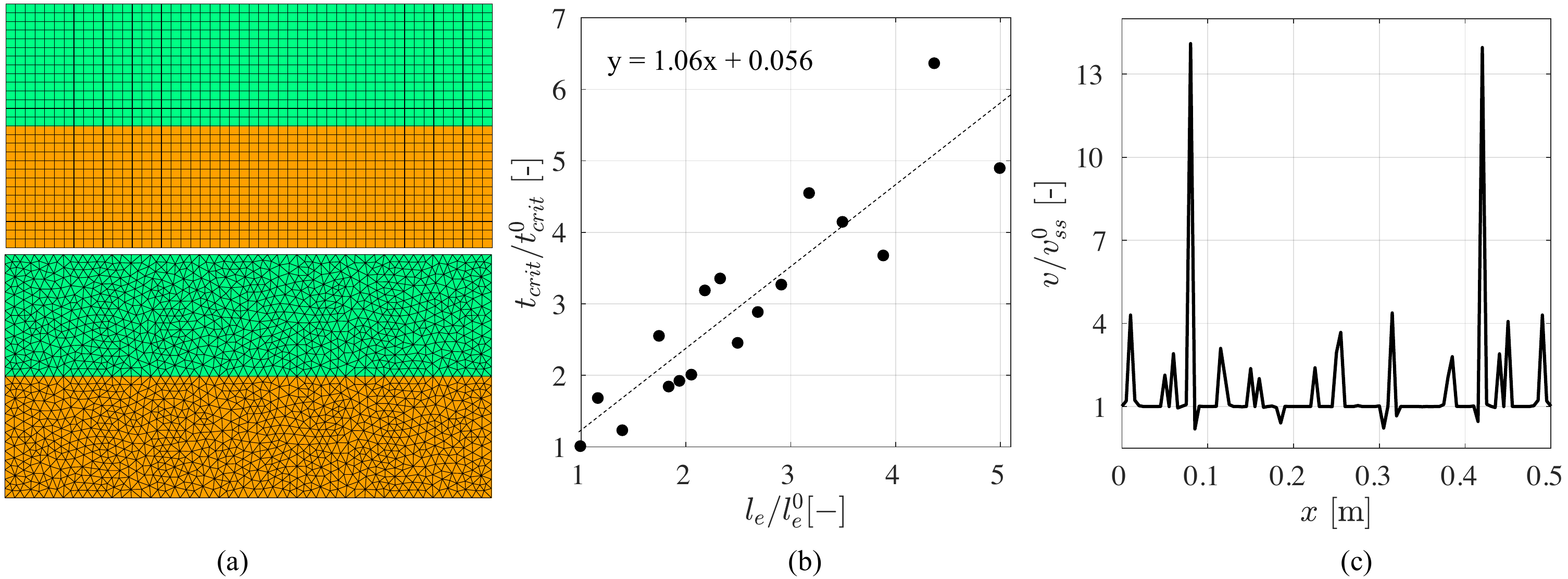}
	\caption{(a) Finite element representation of the structured and unstructured mesh of the blocks by discretizing the interface with 50 elements. (b) For structured finite element mesh with $\alpha=0.2$, normalized time to start instability is plotted versus normalized finite element mesh size. $l_e^0 = 1.42$ [mm] for which the interface is discretized by 350 elements. $t_{crit}^0$ is the duration of steady state sliding before instability starts for the case of $l_e^0$. For the coarsest case $l_e/l_e^0 = 5$, the interface is discretized by 70 elements. (c) Normalized velocity over the interface at the initiation of instability for the case of unstructured mesh with 100 elements on the interface. }
	\label{instability-1}
\end{figure}

We have investigated the same problem using a boundary integral formulation of the elastodynamic equations, Equation \ref{motion-1}. In this method, which is denoted as BIM hereafter, only the interface is simulated and discretized, and the elastic bulk is considered through influence functions, check the Reference \cite{geubelle_spectral_1995} for more details.  We have observed that the required time step to maintain steady state sliding is one order of magnitude higher compared to the finite element simulation. To investigate the origin of this small time step requirement, we have used different finite element mesh sizes to discretize the elastic blocks. In Figure \ref{instability-1}a (top), a generic uniform structured mesh of the blocks is depicted. For the case of coarsest mesh, the interface is discretized by 70 elements, and it is discretized by 350 elements for the finest mesh. For all cases, we choose $\alpha=0.2$. In all simulations, the problem starts with steady state sliding of the interface for a certain period of time, and then the oscillation in interfacial slip velocity starts. The amount of time taken for the instability to start is denoted as $t_{\text{crit}}$, and it is denoted as $t^0_{\text{crit}}$ for the case of the finest mesh. The instability mode for all cases is similar to the plot shown in Figure \ref{250-instability}b, in which the slip velocity at every interfacial node pairs oscillates uniformly. In Figure \ref{instability-1}b, the normalized time to instability $t_{\text{crit}}/t^0_{\text{crit}}$ is plotted versus normalized finite element mesh size $l_e/l_e^0$. $l_e^0$ is the element size corresponding to the finest mesh, which is equal to $500/350 = 1.42$ [mm]. The linear fit of the data indicates that the normalized time to instability linearly increases with the normalized finite element size in an average sense. We infer that the instability starts as the numerical noise in the explicit algorithm travels back to the interface from the first row of bulk internal nodes located just above or below the interface. This conclusion is further confirmed by simulating the same test using solid blocks discretized by unstructured triangular finite element meshes, see Figure \ref{instability-1}a (bottom) for a generic mesh. Normalized slip velocity on the interface is plotted versus space at the initial stage of instability growth in Figure \ref{instability-1}c. The mode of instability differs from the case of structured meshes, in which the whole interface vibrates uniformly, while in unstructured meshes the positions of nodes located near the interface vary randomly triggering instabilities at seemingly random locations. Note that this source of perturbation is not present in spectral (BIM) approaches as there is no internal discretization, leading to time steps an order of magnitude higher. Finally, to increase computational efficiency along with numerical accuracy, more accurate time integration schemes such as higher order Runge-Kutta method with adaptive time stepping could be employed.

\section{Frictional rupture propagation}\label{rupture-propagation}
Next, the state variable at the frictional contact interface is perturbed giving rise to slip velocity localization and subsequent shear rupture propagation. The influence of the finite size of the solid blocks on the long term frictional sliding is presented. An in-depth comparison of the propagating frictional rupture is performed between the pure velocity weakening and the velocity weakening-strengthening friction laws.

\subsection{State variable perturbation analysis using VWS friction law}\label{state-perturbation}
\subsubsection{Without boundary reflection}\label{no-reflection-effect}
We impose an initial steady-state slip velocity of $v^0_{ss} = 3.8931 \times 10^{-4}$ [m/s], corresonding to the purple star shown by the dashed arrow in Figure \ref{elas-elas-setup-curve}b. This point is located on the unstable weakening branch, which can lead to shear rupture propagation on the interface. At any interfacial node pair, the initial relative slip velocity is equal to $v^0_{ss}$, and the state variable is constant $\phi_{ss} = D/v^0_{ss}$. Next, we create a small perturbation of the state variable $\phi(x) = \phi_{ss}$ over the interface in the form of a sinus function: $\phi(x) = \phi_{ss} + \varepsilon \sin(2\pi x/L)$, with $\varepsilon = 10^{-4}$, and $x$ the horizontal coordinate with origin at the left corner of the interface, see Figure \ref{elas-elas-setup-curve}a. In this example, we choose the interface length $L = 0.5$ [m] and the height of each solid block $H = 1.25$ [m]. $H$ is large enough so that the waves reflected back from the top and bottom boundaries do not reach the interface to alter the rupture propagation within the allocated simulation time. The interface is discretized by 250 elements, and the same finite element size is used to mesh uniformly  the two solids with linear quadrilateral elements. The time step is set to $2\%$ of the stable time step determined by the CFL condition, and the total simulation time is 3 [ms]. The numerical example considered in this section is also solved by BIM and reported by Barras \cite{Barras_Thesis}. To validate the proposed finite element framework of this paper, the numerical results will be compared to the numerical solution obtained by BIM.
\begin{figure}[t!]
	\centering 
	\includegraphics[width=\textwidth]{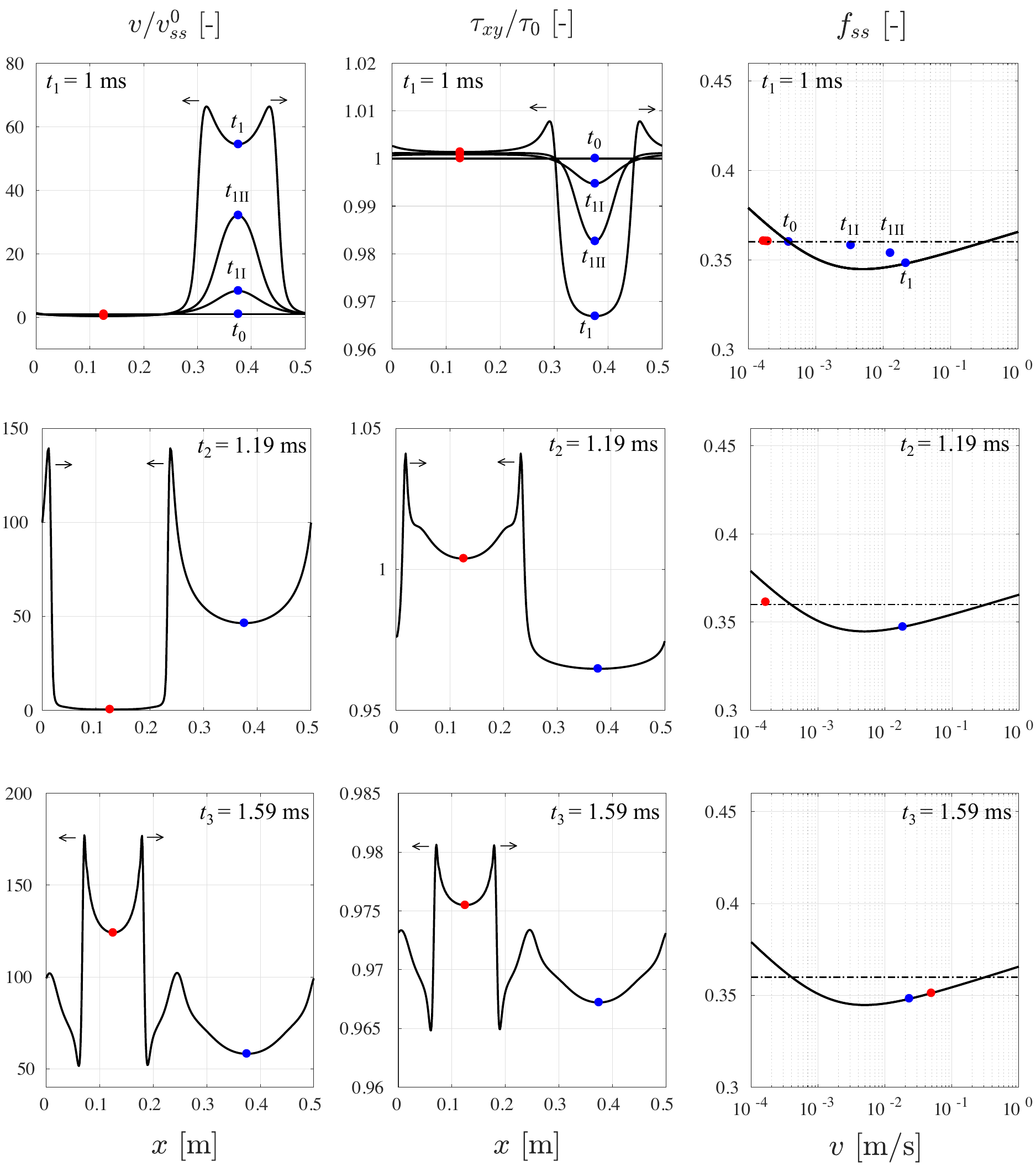}
	\caption{In columns 1 and 2, respectively, evolution of the interfacial slip velocity normalized by $v^0_{ss}$ and interfacial shear traction normalized by $\tau_0$ are presented at different time instants. In the third column, friction coefficient versus slip velocity at $x = 0.125$ [m] and $x = 0.375$ [m] on the interface, marked as red and blue points in columns 1 and 2, are shown with respect to the steady-state friction curve (continuous black curve).}
	\label{vel-evolution-all}
\end{figure}

\begin{figure}[t!]
        \centering
        \begin{subfigure}[b]{0.45\textwidth}
                \centering
                \includegraphics[width=\textwidth]{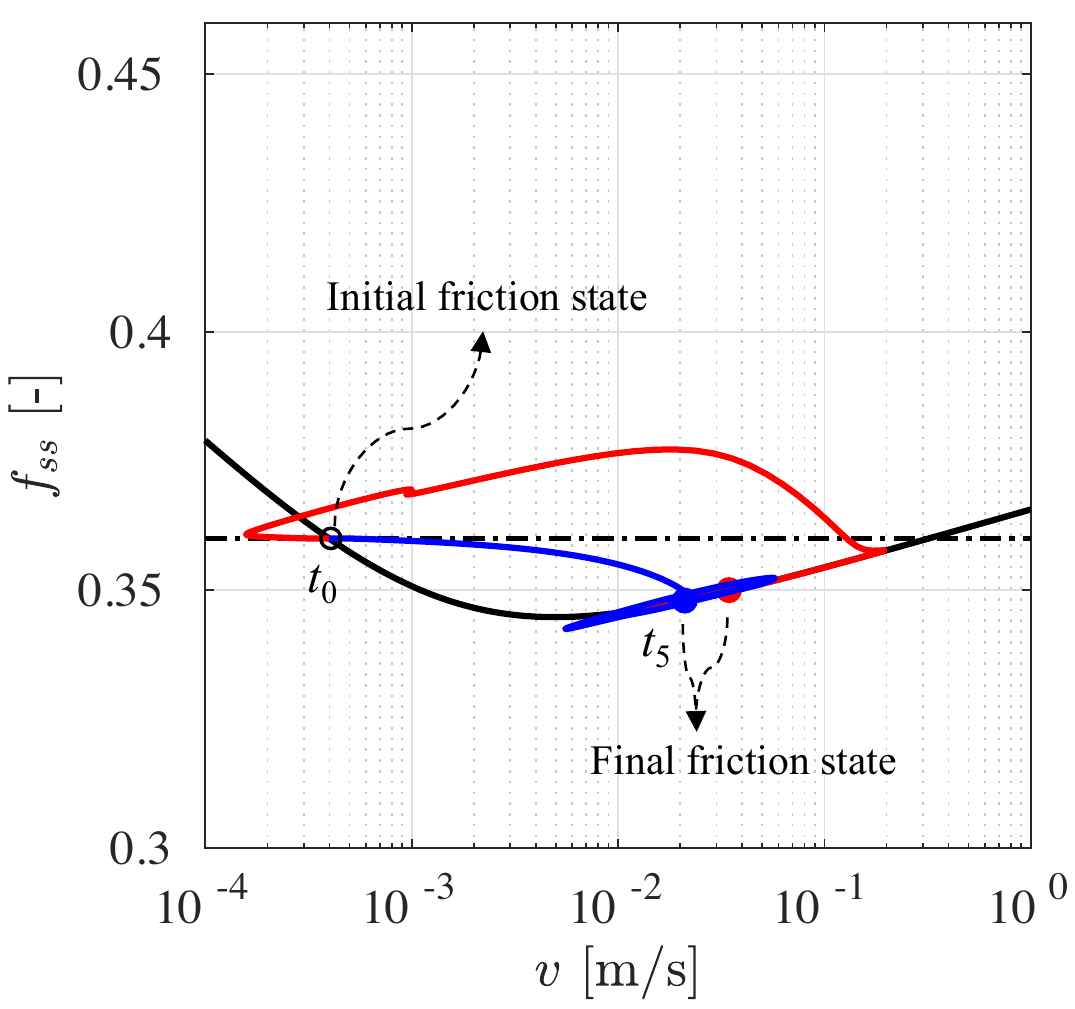}
                \caption{}
        \end{subfigure}
        \begin{subfigure}[b]{0.45\textwidth}
                \centering
                \includegraphics[width=\textwidth]{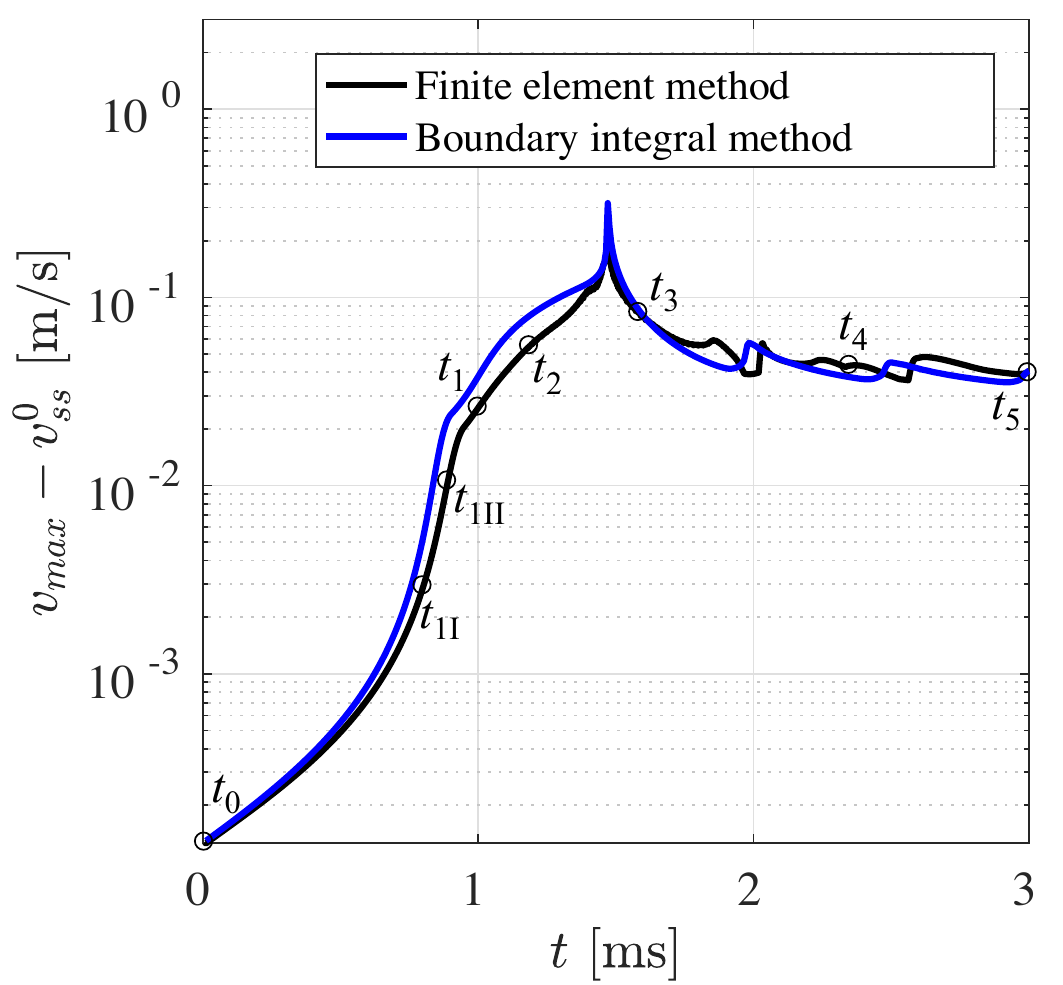}
                \caption{}
        \end{subfigure}
        \caption{(a) Friction coefficient evolution at two points on the interface located at $x = 0.125$ [m] and $x = 0.375$ [m] is plotted in red and blue, respectively. The dash-dotted line is the far field applied boundary condition. Notice that without wave reflections from outer boundaries, the frictional forces at the interface end up at a level below the far field shear stress. The blocks are thus not in macroscopic equilibrium. (b) Variation of the maximum slip velocity on the interface minus the initial steady state velocity during the simulation time for FEM and BIM simulations. The velocity reaches a temporary steady state with the FEM simulation, similar to the one obtained with the BIM simulation which assumes an infinite domain.}
        \label{points-vmax}
\end{figure}

The evolution of slip velocity and shear traction on the interface at different times during the simulation are depicted in the first and second columns of Figure \ref{vel-evolution-all}. The slip velocity is normalized by the initial steady state slip velocity $v^0_{ss} = 3.8931 \times 10^{-4}$ [m/s], and the interfacial shear traction is normalized by the far field applied shear traction $\tau_0 = 0.36$ [MPa]. To shed light on the evolution of friction coefficient during the rupture propagation, two nodes marked with red and blue circles are selected on the interface at quarter and three-quarter of $L$, which are also the location of the maximum and minimum peaks of the sinus perturbation function, respectively. This evolution is illustrated in the third column of Figure \ref{vel-evolution-all} at the same time instants as in the first and second columns. Initially, the whole interface is under steady state sliding condition, $t=t_0$ in the first row of Figure \ref{vel-evolution-all}, at which $v/v^0_{ss} = \tau_{xy}/\tau_0 = 1$, and both blue and red circles are on the weakening branch of $f_{ss}$. When the sinus form perturbation in state variable is applied on the interface, three subsequent distinct phases are recognizable. These three phases are illustrated in the three rows of Figure \ref{vel-evolution-all}. First, the slip velocity localizes and grows on the right half of the interface, where the state variable is decreased due to perturbation. As seen in the first row of Figure \ref{vel-evolution-all}, between $t_0 = 0$ to $t_1 = 1$ [ms] (with intermediate steps $t_{1I}$ = 0.801 [ms] and $t_{1II}$ = 0.9 [ms]), the growth in slip velocity in the right half of the interface is simultaneous to the decrease of the shear traction. During this phase, the coefficient of friction at the blue point moves from the weakening branch to the strengthening one. On the left half of the interface where the state variable is increased, the slip velocity decreases slightly, and the shear traction increases by a small amount. Finally, localization and growth in slip velocity leads to nucleation of two shear rupture fronts, visible in the curve corresponding to $t=t_1$. In the second phase, the two rupture fronts start to propagate in opposite directions, and a snapshot of this phase is plotted in the second row of Figure \ref{vel-evolution-all} at $t_2=1.19$ [ms]. The two small arrows on the plots of first and second columns in Figure \ref{vel-evolution-all} represent the propagation direction of the two rupture fronts. At $t=t_2$, the rupture fronts have not yet arrived to the position of the red circle, where the slip velocity is negligible compared to the rest of domain. Due to periodic boundary conditions, the two rupture fronts finally collide and pass each other. The third row of Figure \ref{vel-evolution-all} represents this phase at $t_3 = 1.59$ [ms], right after the two fronts intersect. After the rupture fronts arrive to the position of the red point, the friction coefficient at this point moves to the strengthening branch of the steady state curve shown in the third column of Figure \ref{vel-evolution-all} at $t_3 = 1.59$.

\begin{figure}[t!]
	\centering 
	\includegraphics[width=\textwidth]{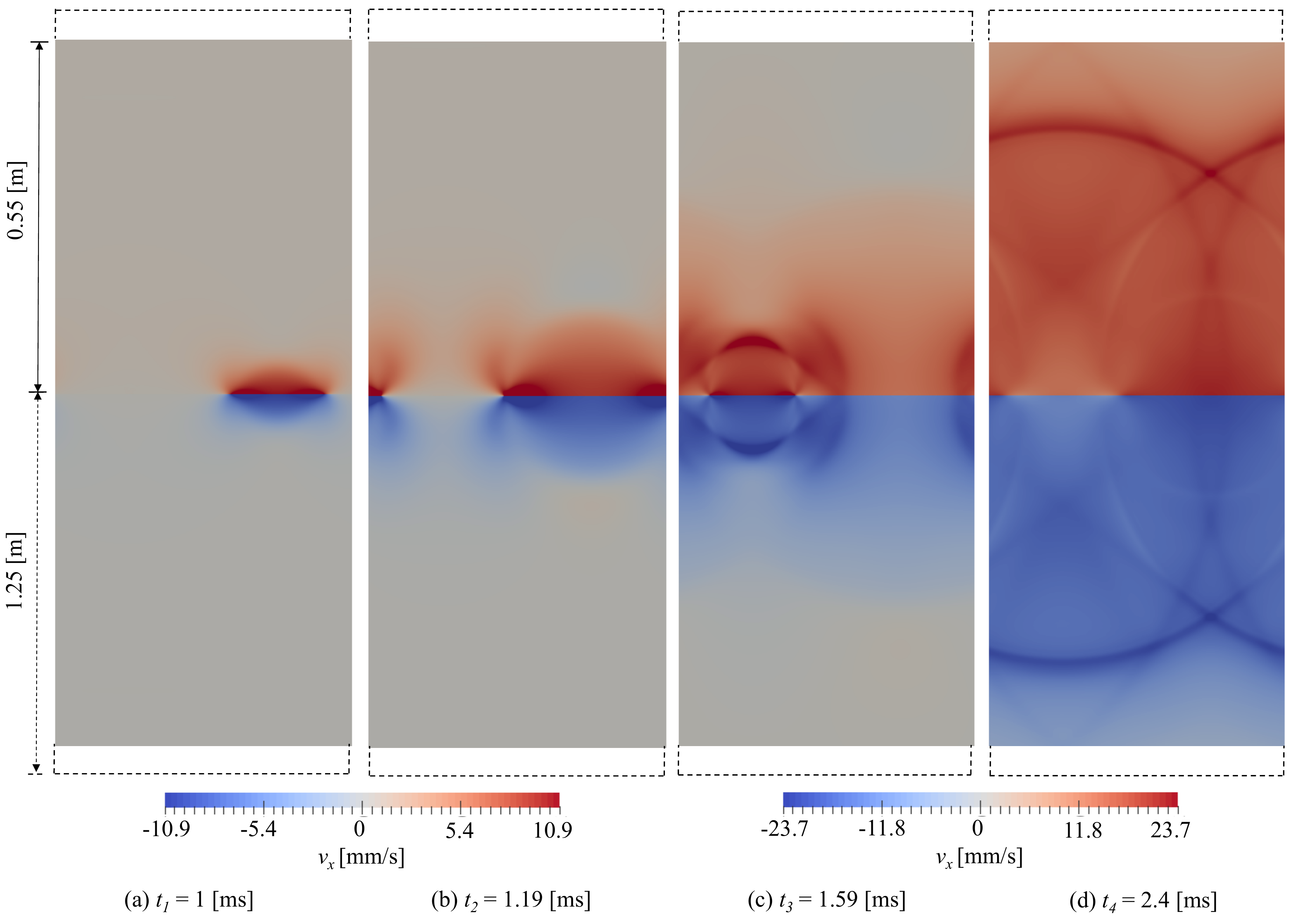}
	\caption{Contours of horizontal component of velocity field at four time instants during the simulation. Solid blocks shown in this figure are cut at the height of 0.55 [m] for better visualization, while their height in the simulation is 1.25 [m].}
	\label{inifinite-vx-contour}
\end{figure}

The full evolution of the friction coefficient during the simulation at the blue and red points positions is depicted in Figure \ref{points-vmax}a. In the beginning, they are both placed at the initial steady state position as marked on the plot, and they move to the strengthening branch as the rupture fronts travel through the interface. The final coefficient of friction at these two points is also shown on the plot at $t=t_5=3$ [ms]. Note that this final value is not in equilibrium with the far field loading, marked as purple triangle in Figure \ref{elas-elas-setup-curve}b. This issue will be discussed further in the next section. The maximum slip velocity at the interface minus the initial steady state velocity, $v_{max}-v^0_{ss}$, is plotted on logarithmic axis versus time in Figure \ref{points-vmax}b. Three distinct phases are easily distinguishable on this plot. After a rapid growth of slip velocity localization, from $t_0$ to after $t_{1II}$, a sudden change in the slip velocity rate becomes visible. $t_1$ to $t_2$ corresponds to the rupture propagation phase before the two fronts meet. The spike in maximum velocity observed between $t_2$ and $t_3$ is the moment that the two fronts meet, and the subsequent steady state corresponds to several propagation of the fronts accross the interface. Since in this simulation wave reflections from the top and bottom boundaries do not affect the interfacial response, the finite element solution can be compared to the results of BIM which assumes an infinite domain \cite{Barras_Thesis}. The FEM and BIM results are in good agreement, Figure \ref{points-vmax}b.

Contours of the horizontal component of velocity field, $v_x$, are plotted in Figure \ref{inifinite-vx-contour} at four different time instants: $t_1$ and $t_2$ during the propagation phase, and $t_3$ and $t_4$ after the two fronts meet and cross. Note that the real height of each block is 1.25 [m], and only 0.55 [m] is shown for clarity. 


\begin{figure}[t!]
        \centering
        \begin{subfigure}[b]{0.45\textwidth}
                \centering
                \includegraphics[width=\textwidth]{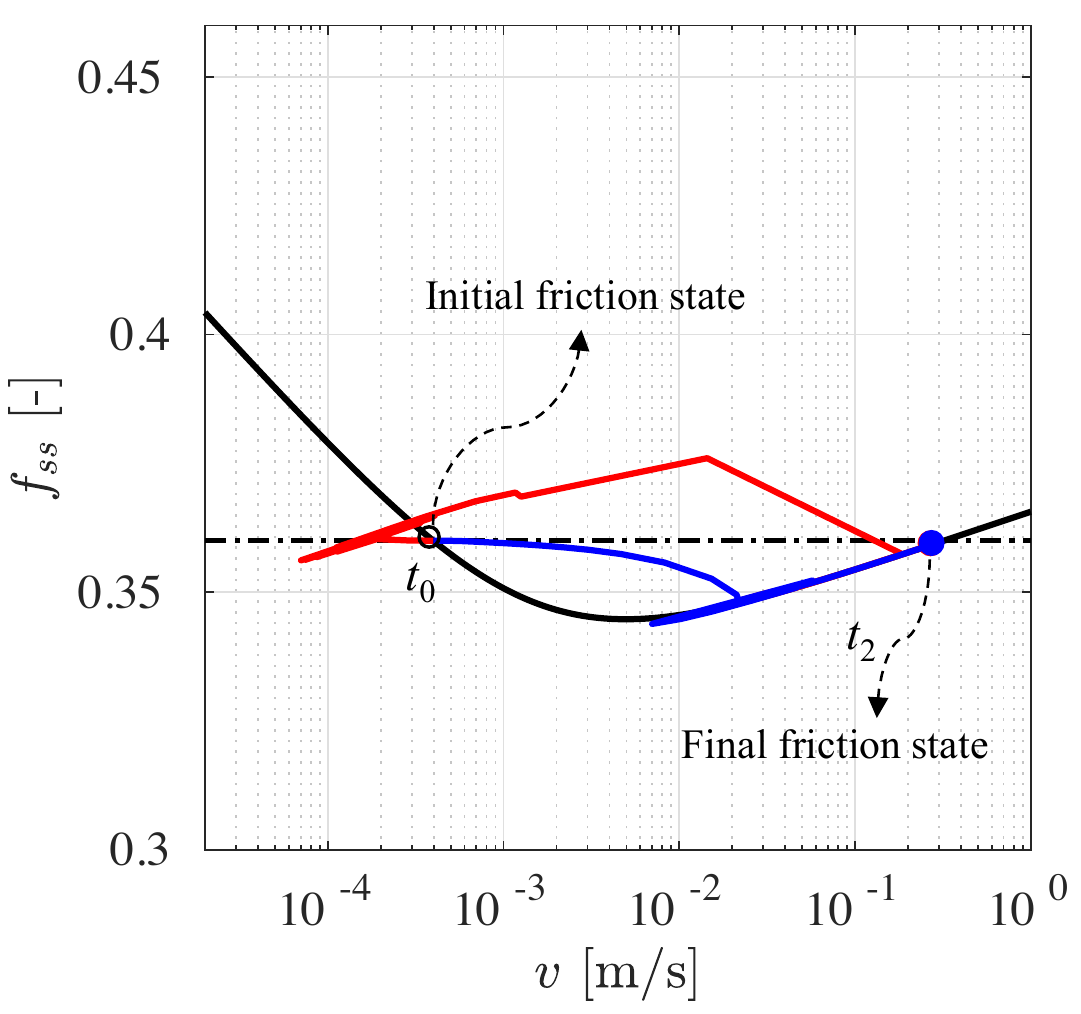}
                \caption{}
        \end{subfigure}
        \begin{subfigure}[b]{0.45\textwidth}
                \centering
                \includegraphics[width=\textwidth]{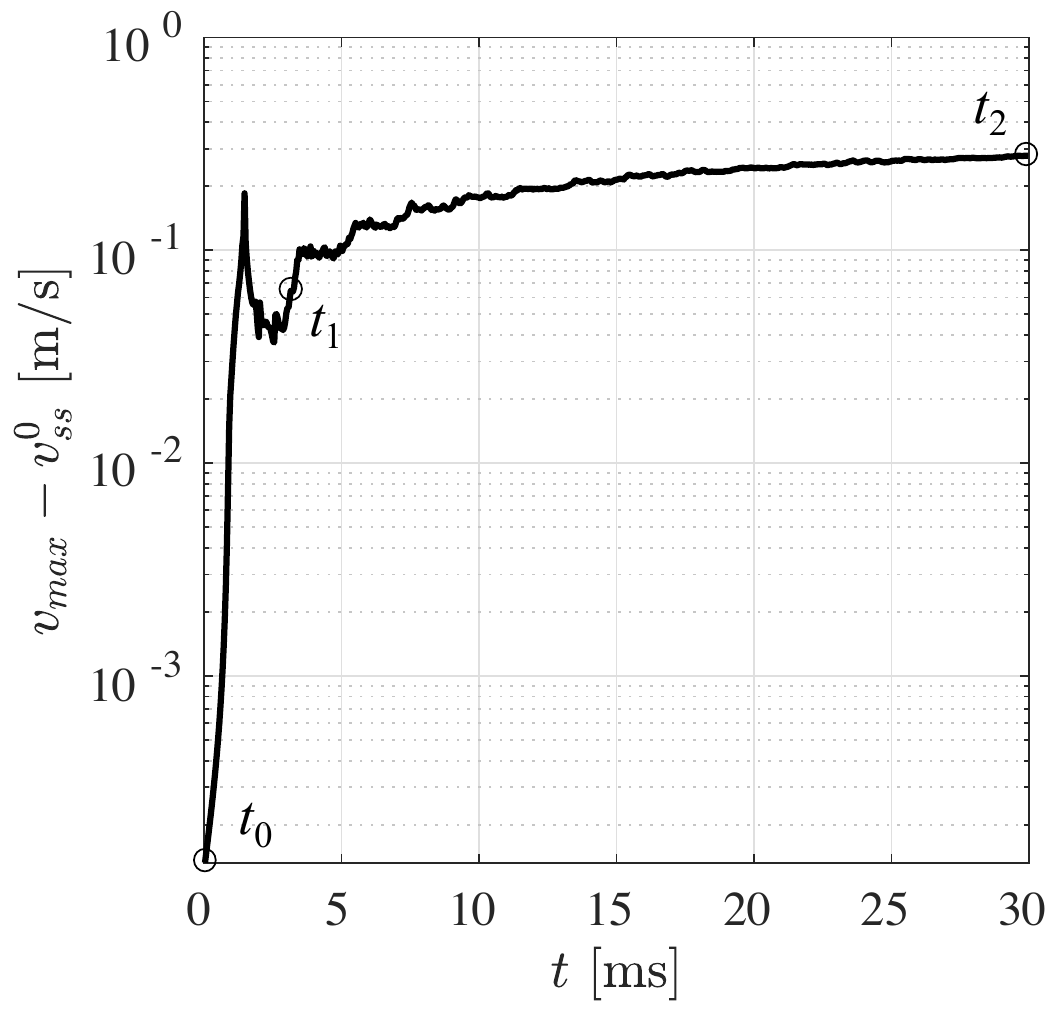}
                \caption{}
        \end{subfigure}
        \caption{(a) Friction coefficient evolution at two points on the interface at $x = 0.125$ [m] and $x = 0.375$ [m] plotted in red and blue, respectively. The dash-dotted line is the far field applied boundary condition. Notice that, at the final simulation time, after multiple wave reflections, the frictional forces at the interface roughly equilibrate the far field shear stress. Macroscopic equilibrium is achieved and the blocks are moving at a constant velocity. A permanent frictional sliding steady state is reached with the VWS friction law. (b) Variation of maximum slip velocity on the interface minus the initial steady state velocity during the simulation time.}
        \label{points-vmax-finite}
\end{figure}

\subsubsection{Effect of boundary reflection}\label{reflection-effect}
\begin{figure}[t!]
	\centering 
	\includegraphics[width=1\textwidth]{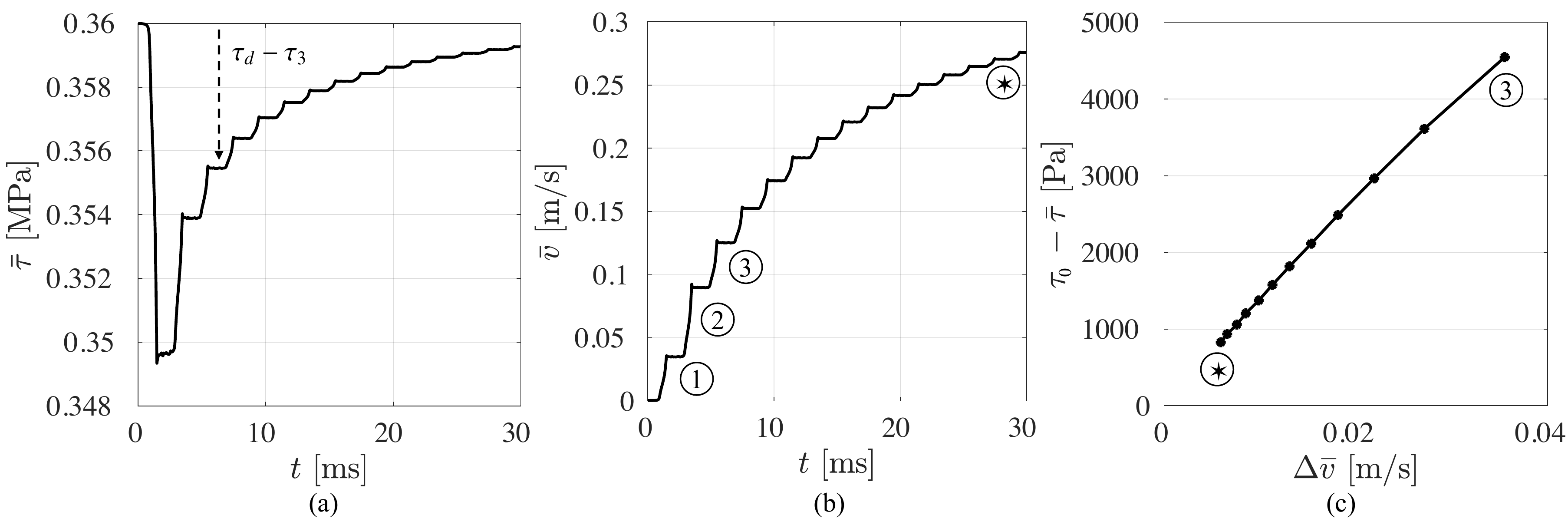}
	\caption{(a) Variation of average shear traction on the interface versus time. (b) Variation of average slip velocity on the interface versus time. (c) Average stress drop versus average velocity jump, corresponding to each times boundary-reflected waves reach back the interface.}
	\label{finite-reflections}
\end{figure}

The numerical example presented in Section \ref{no-reflection-effect} is now revisited to investigate the finite boundary effect on the interfacial response. We emphasize again that this cannot be investigated with infinite domain methods such as BIM. In this example, the height of the two blocks is shortened to $H=0.5$ [m], and the total time of the simulation is set to $30$ [ms]. Therefore, the waves reflected from the top and bottom boundaries will travel back to the interface and govern the evolution of the frictional response. In Figure \ref{points-vmax-finite}a, the evolution of the friction coefficient at the red and blue points are presented again. Clearly both points gradually converge to the equilibrium position, which is the intersection of $f_{ss} = 0.36$ (dash-dotted line) and the steady state curve illustrated in Figure \ref{elas-elas-setup-curve}b as purple triangle. $v_{max}-v^0_{ss}$ versus time is plotted in Figure \ref{points-vmax-finite}b. The same trend observed in Figure \ref{points-vmax}b is also observed in the initial part of Figure \ref{points-vmax-finite}b, which is followed by sudden jump in the value of maximum velocity at $t_1$. This is the time when the reflected waves from the top and bottom boundaries reach the interface and escalate the slip velocity. Such jumps are repeated after each wave reflections until the slip velocity of the interface generates an interfacial shear traction that is in equilibrium with the far field shear loading. In Figure \ref{finite-reflections}a and b, shear tractions and slip velocities are averaged over the interface and plotted versus time, respectively. The first jump in $\bar{v}$ in Figure \ref{finite-reflections}b corresponds to the sudden decrease of $\bar{\tau}$ in Figure \ref{finite-reflections}a, which represents the nucleation and propagation phases of the frictional shear rupture before arrival of the reflected waves. The second jump corresponds to the first time that the boundary reflected waves arrive to the interface, when the propagation of shear traction and slip velocity singularities are disturbed. The following jumps take place subsequently each time the reflected waves impact the interface and increase the average slip velocity, resulting in a gradual convergence of $\bar{\tau}$ to the far field loading $\tau_0$. In Figure \ref{finite-reflections}c, $\tau_0-\bar{\tau}$ is plotted versus the average velocity jump, considered from the second time that the reflected waves impact the interface, denoted by "3" in Figure \ref{finite-reflections}b. This choice was made because after the first wave reflection, represented by "2" on Figure \ref{finite-reflections}b, some trace of singularities are still observed in $\tau$ and $v$ profiles. These profiles become close to uniform over the interface after the following incident wave, and their average values are representative of the interfacial quantities. Note that the shear stress drop, $\tau_0-\bar{\tau}$, decreases towards zero as a linear function of the slip velocity jump. Theoretical arguments \cite{Barras_partI} can be made to relate these jumps to order of $\mu/2c_s$, and will be investigated further in the future.

\begin{figure}[t!]
	\centering 
	\includegraphics[width=0.7\textwidth]{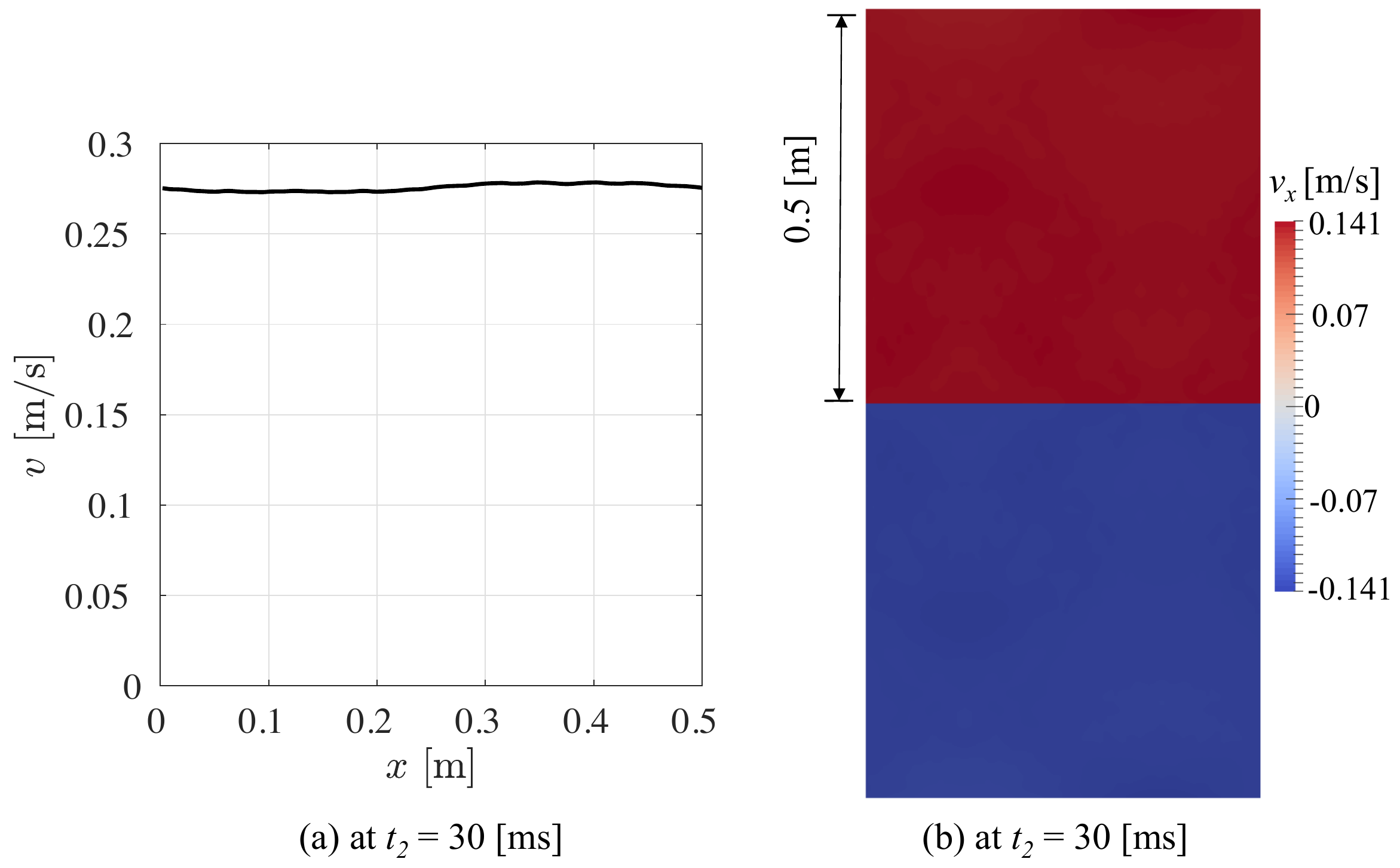}
	\caption{(a) Slip velocity at the interface, and (b) contour of $v_x$ in the two solid blocks at the end of the simulation with the effect of boundary reflections, showing that the two blocks move uniformly in opposite directions.}
	\label{vx-finite-final}
\end{figure}

\begin{figure}[t!]
	\centering 
	\includegraphics[width=\textwidth]{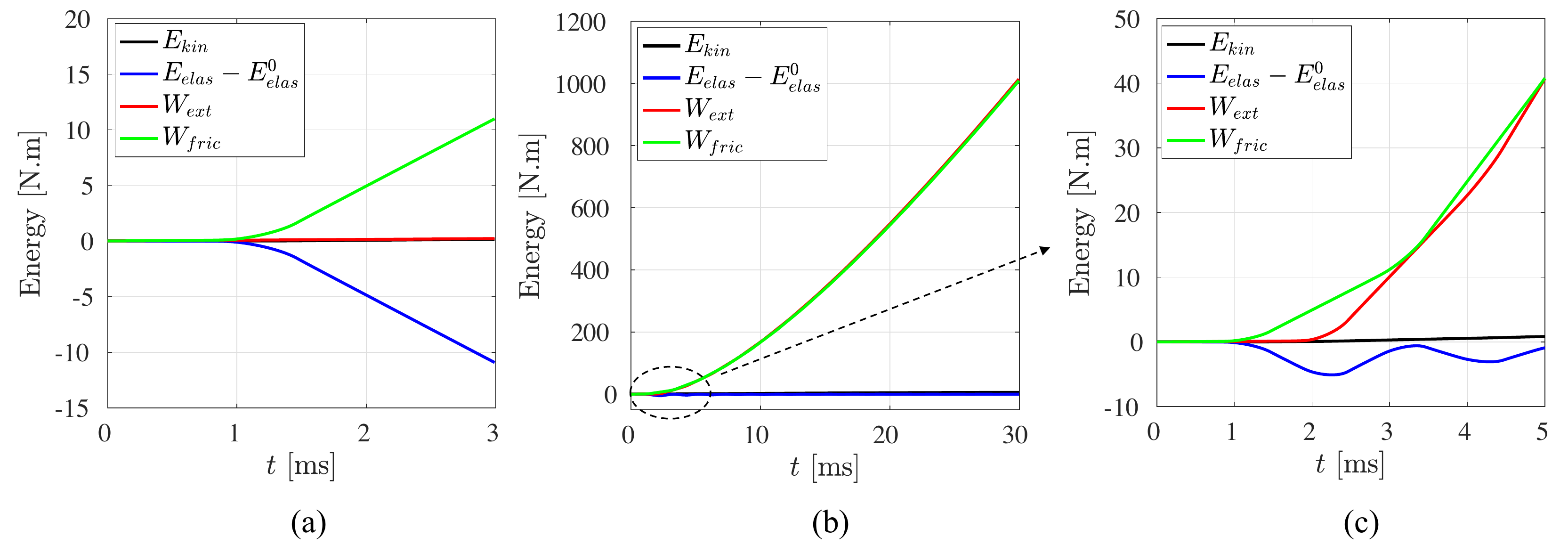}
	\caption{Evolution of kinetic energy $E_{\text{kin}}$, elastic energy $E_{\text{elas}}$ (minus initial value due to far field loading), external work $W_{\text{ext}}$, and frictional work $W_{\text{fric}}$ versus time for the simulation (a) with no boundary wave reflection and (b) with boundary wave reflections. (c) Zoomed view of subplot (b) at the time when reflected waves reach the interface.}
	\label{energy-comparison}
\end{figure}

The slip velocity at the interface at the end of the simulation $t_2 = 30$ [ms] is plotted in Figure \ref{vx-finite-final}a. One can see that the whole interface slides with the slip rate corresponding to the ``final friction state" marked on Figure \ref{points-vmax-finite}a, which results in 0.36 MPa interfacial shear stress equilibrating the applied far field shear traction. Contours of $v_x$ are presented in Figure \ref{vx-finite-final}b, which shows the whole top and bottom blocks slide with uniform velocity throughout the bulk. It should be noted that the shear traction and slip velocity singularities observed during the rupture propagation phase, demonstrated in Figure \ref{vel-evolution-all}, vanish when the reflected waves from the top and bottom boundaries reach the interface, and the whole interface slides with approximately uniform slip velocity.

Finally, the evolution of different energy quantities, kinetic energy $E_{\text{kin}}$, elastic energy $E_{\text{elas}}$, external work $W_{\text{ext}}$, and frictional work $W_{\text{fric}}$, are compared for the two cases, with and without wave reflections. Each of these quantities at any time $t_n$ are calculated as
\begin{equation}\label{E_kin}
E_{\text{kin}} = \frac{1}{2} \int_{\Omega}\rho\dot{u}_i\dot{u}_i ~\text{d}\Omega = \frac{1}{2} \dot{\mbf u}^{T}\mbf{M}\dot{\mbf u}
\end{equation}
\begin{equation}\label{E_int}
E_{\text{elas}} = \int_{0}^{t} \bigg[\int_{\Omega}\sigma_{ij} \dot{\varepsilon}_{ij} ~\text{d}\Omega \bigg] \text{d}t = \frac{1}{2} \int_{\Omega} \boldsymbol{\sigma\varepsilon} ~\text{d}\Omega
\end{equation}
\begin{equation}\label{W_ext}
W_{\text{ext}} = \int_{0}^{t} \bigg[\int_{\Gamma_t}T_{i} \delta\dot{u}_{i} ~\text{d}\Gamma_t \bigg] \text{d}t = \sum_{n=1}^{t_n} \mbf{T} \delta\mbf{\dot{u}}~\Delta t
\end{equation}
\begin{equation}\label{W_fric}
W_{\text{fric}} = \int_{0}^{t} \bigg[\int_{\Gamma_C}R^T_{i} \llbracket \delta\dot{u}_{i} \rrbracket ~\text{d}\Gamma_c \bigg] \text{d}t = \sum_{n=1}^{t_n} \mbf{R}^T \llbracket\delta\mbf{\dot{u}}\rrbracket~\Delta t
\end{equation}

In all equations above, the time subscript is dropped for clarity. Evolution of these quantities is presented in Figure \ref{energy-comparison}.  In Figure \ref{energy-comparison}a, energy variations with time are plotted for the case without wave reflections from boundaries. After approximately 1 [ms] the propagation of slip fronts begins. The friction work then increases, and the stored elastic energy decreases by the same amount. Kinetic energy and external work are negligible quantities on this time scale. In Figure \ref{energy-comparison}b, the same energy terms are plotted but accounting wave reflections at boundaries, and a zoomed view of the first 5 [ms] is plotted in Figure \ref{energy-comparison}c. The same trend is observed in the first 2 milliseconds until the emitted waves from the interface reach the top and bottom boundaries and travel back to the interface. Approximately all external work is converted to frictional work. Although blocks move at a finite velocity, the kinetic energy remains small. Elastic energy is also negligible, elastic vibrations are progressively dying out due to friction at the interface which acts as an effective viscosity.

\begin{figure}[t!]
	\centering 
	\includegraphics[width=\textwidth]{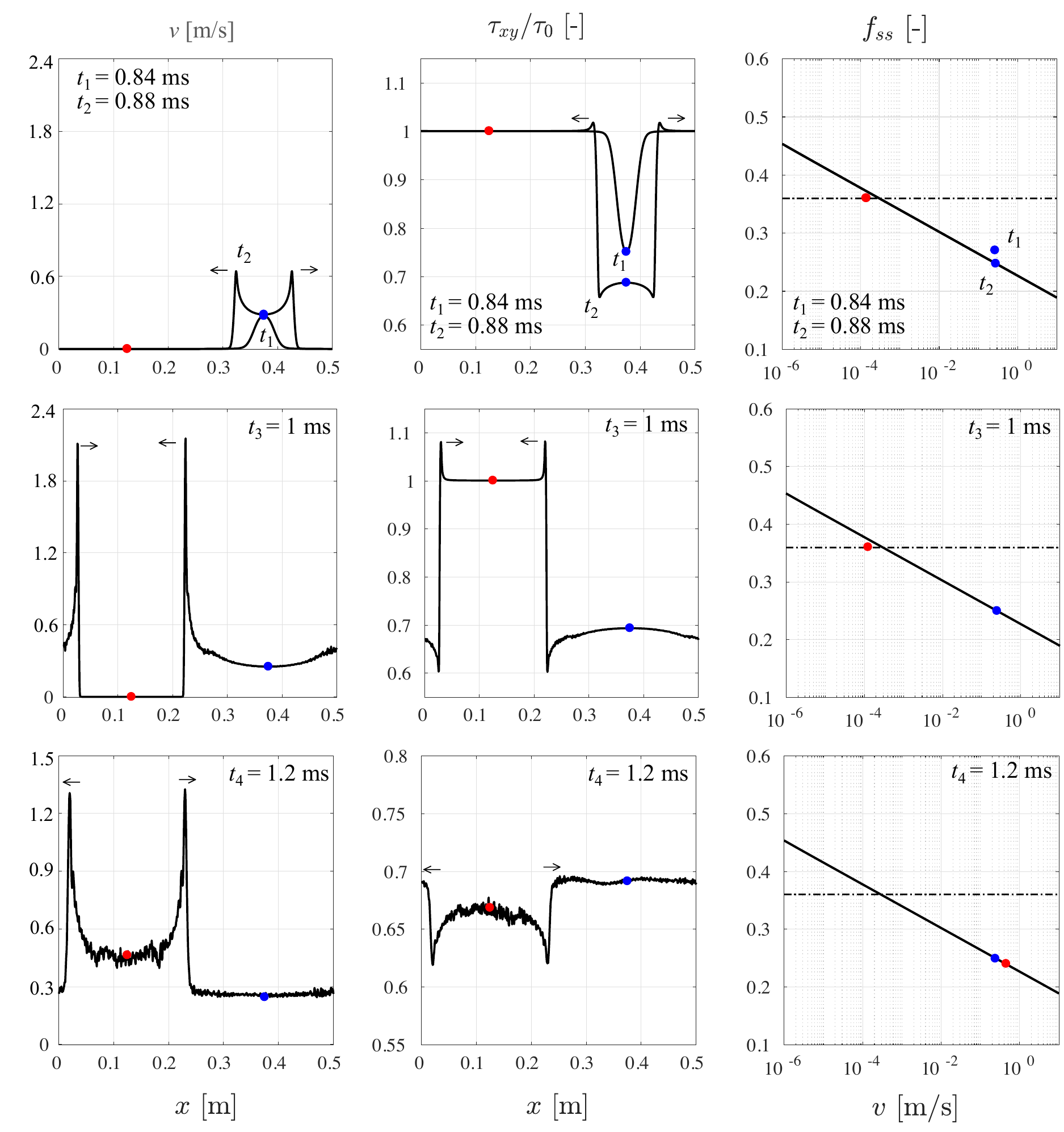}
	\caption{In columns 1 and 2, respectively, evolution of the interfacial slip velocity and interfacial shear traction normalized by $\tau_0$ are presented at different time instants. In the third column, friction coefficient versus slip velocity at $x = 0.125$ [m] and $x = 0.375$ [m] on the interface, marked in red and blue in columns 1 and 2, is shown with respect to the steady-state curve.}
	\label{VW-vel-evolution}
\end{figure}

\begin{figure}[h!]
	\centering 
	\includegraphics[width=0.8\textwidth]{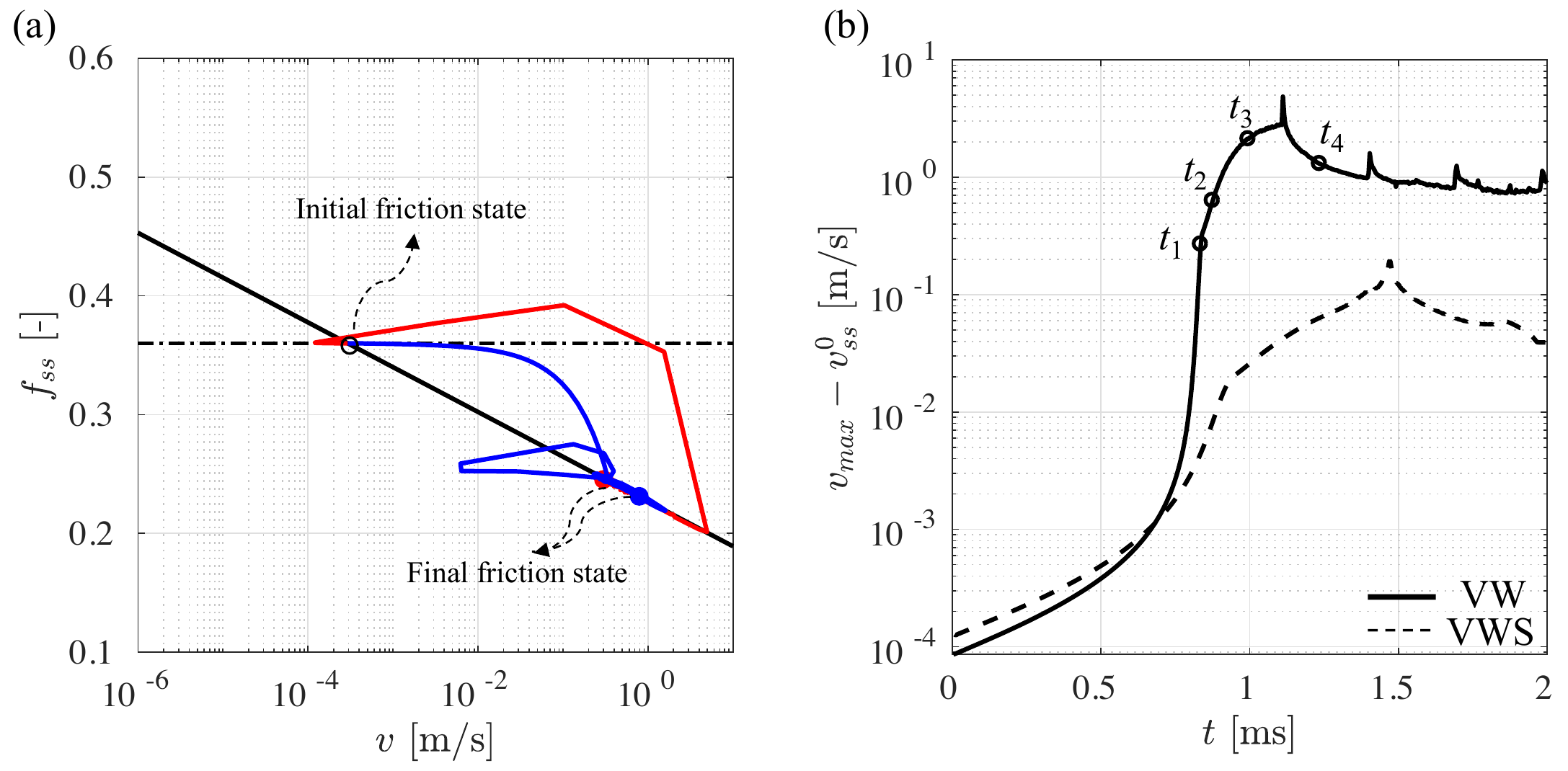}
	\caption{(a) Friction coefficient evolution at two points on the interface located at $x = 0.125$ [m] and $x = 0.375$ [m] plotted in red and blue, respectively. (b) Variation of maximum slip velocity on the interface minus the initial steady state velocity during the simulation time.}
	\label{VW-points-vmax}
\end{figure}

\begin{figure}[h!]
	\centering 
	\includegraphics[width=1\textwidth]{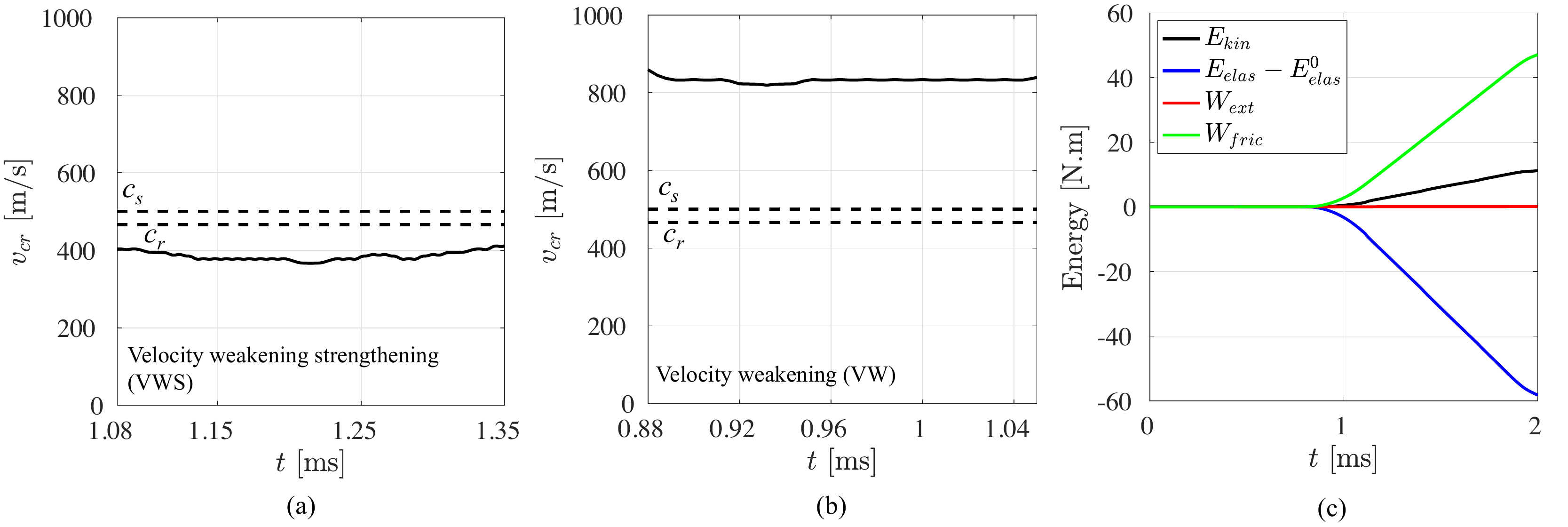}
	\caption{Propagation velocity of the left moving shear rupture front is recorded for both cases of velocity weakening-strengthening and pure velocity weakening and are plotted in (a) and (b), respectively. $c_s$ and $c_r$ are the shear and Rayleigh wave speeds. (c) Evolution of different energy quantities for the example using VW law. $E_{elas}$ is plotted after subtracting the initial stored elastic energy due to the far field loading.}
	\label{Vcr-energy}
\end{figure}

\begin{figure}[t!]
	\centering 
	\includegraphics[width=0.65\textwidth]{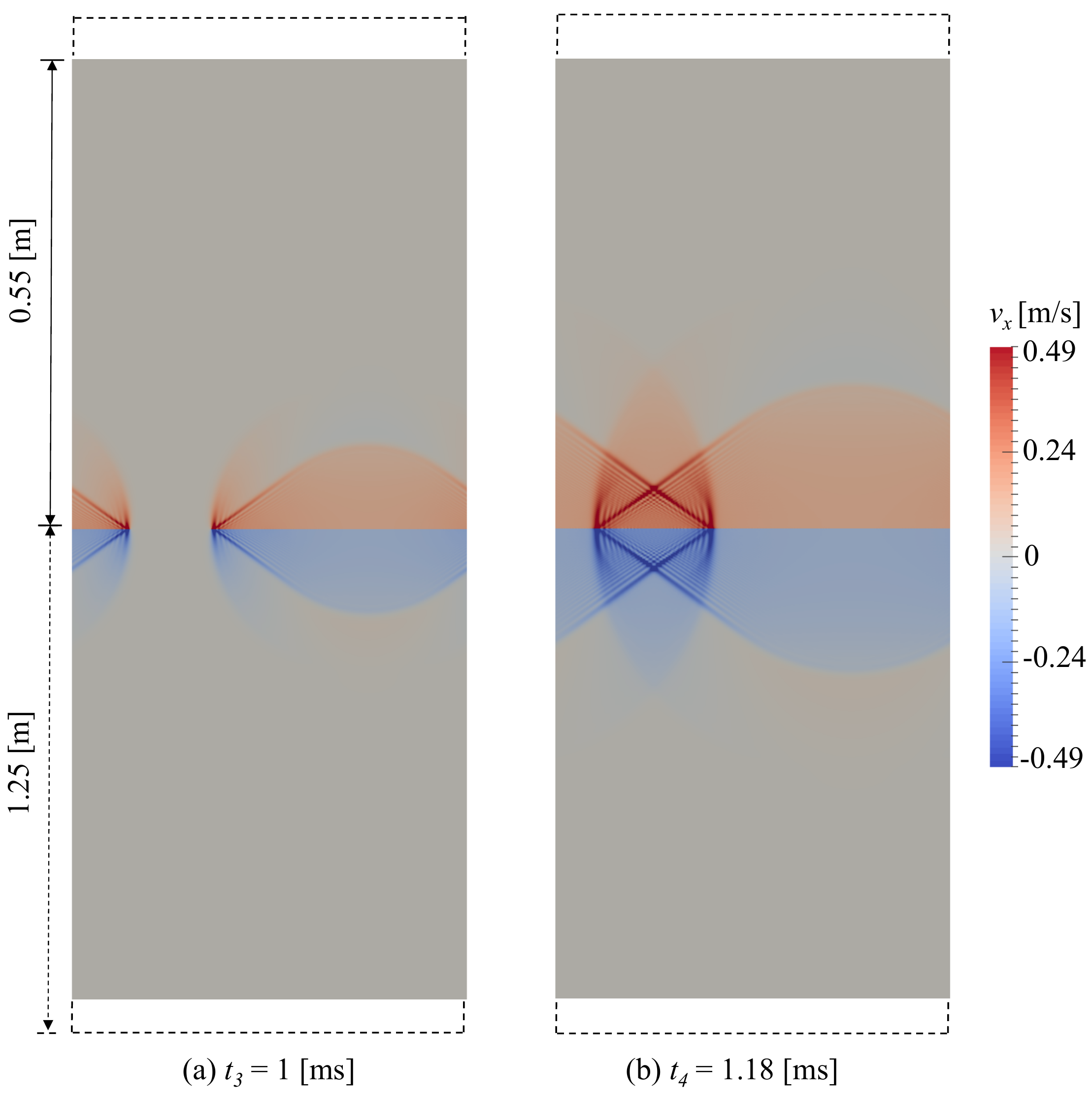}
	\caption{Contour of horizontal component of velocity field at two time instants during the initial propagation phase of shear rupture fronts and after they meet and cross each other. Solid blocks shown in this figure are cut at the height of 0.55 [m] for clarity, while their height in the simulation is 1.25 [m].}
	\label{VW-Vx-contour}
\end{figure}

\subsection{Perturbation analysis using VW law}\label{perturbation-VW}
Finally we investigate the same perturbation problem of Section \ref{state-perturbation} but this time using the pure velocity weakening law, see Figure \ref{elas-elas-setup-curve}b blue curve. $v^0_{ss}$ corresponding to the friction coefficient of 0.36 is slightly less than the one calculated for the VWS curve and is equal to $2.93627\times10^{-4}$ [m/s]. Boundary and initial conditions explained in Section \ref{state-perturbation} are used, and the same numerical procedure is repeated. 

The three phases of slip velocity localization and growth, nucleation and propagation of the shear rupture fronts, and post-crossing propagation of the two rupture fronts, which were reported in Section \ref{state-perturbation} for VWS case are also observed here. These phases are plotted in Figure \ref{VW-vel-evolution}, and the structure of the plots are the same as the one explained for Figure \ref{vel-evolution-all}. As the slip velocity localizes on the interface, the coefficient of friction at the position of the blue point shifts quickly to another location on the steady state curve with higher value of slip velocity and lower coefficient of friction. The red point also joins that location after the rupture front reaches its location. In recent work, we have shown how the stress drop and slip velocity (i.e. the positions of red and blue points) can be predicted theoretically from radiation damping \cite{Barras_partI}. At this stage, before reflected waves come back to the interface, the system has reached a temporary steady state sliding. The time evolution of the coefficient of friction at those two points is plotted in Figure \ref{VW-points-vmax}a, in which the sharp corners in the red curve are due to the recording frequency of the numerical data. $v_{max}-v^0_{ss}$ versus time is plotted in Figure \ref{VW-points-vmax}b, in which the curve related to the VWS analysis is also plotted for comparison. The slip velocity localization takes place faster and the propagation phase starts earlier for the VW case. In addition, the value of maximum slip velocity when the propagation starts is one order of magnitude higher compared to the VWS analysis. Furthermore, duration of the propagation phase before the two rupture fronts meet is shorter. This implies that, in addition to higher magnitude of slip velocity on the interface, the propagation speed of the fronts is higher. To compare these quantities, the propagation speed of the shear rupture front, denoted as $v_{cr}$, moving leftwards in the second row of Figures \ref{vel-evolution-all} and \ref{VW-vel-evolution}, are measured at every time step during the propagation phase and plotted in Figures \ref{Vcr-energy}a and b for the VWS and VW cases, respectively. $c_s$ and $c_r$ are the shear and Rayleigh wave speeds, which are caluclated as $c_s = \sqrt{E/2\rho(1+\nu)}$ = 500 [m/s] and $c_r = (0.862+1.14\nu)/(1+\nu)\times c_s$ = 466 [m/s]. The shear cracks for the VWS case are in sub-Rayleigh regime, while they are propagating at super shear velocities for the VW problem. This is further evidenced by looking at the $v_{x}$ contour in the two solid blocks at two time instants in the propagation and post-crossing phases, Figure \ref{VW-Vx-contour}. Super shear propagation is confirmed by the conical shape of velocity distribution around the front, which is one of the characteristics of the super shear mode of rupture \cite{shi_properties_2008}. Evolution of different energy quantities are plotted in Figure \ref{Vcr-energy}c. Comparing these quantities with their counterparts obtained from VWS analysis in Figure \ref{energy-comparison}a at 2 [ms], the drop in the stored elastic energy is almost 15 times higher for the case of VW analysis. At this time instant 18$\%$ of this value is converted to kinetic energy, and the rest is dissipated through frictional work.

It must be emphasized that the results presented for the case of VW friction are before return of the reflected waves from boundaries. In Section \ref{reflection-effect} for the VWS case, the boundary-reflected waves lead the blocks to reach global steady state sliding by equilibrating the frictional traction with the far field loading. On the contrary, due to the nature of the VW law presented in Figure \ref{VW-points-vmax}a, it is evident that the reflected waves cause the solid blocks to keep accelerating and a global equilibrium can never be obtained. In the finite element simulation of VW interface, the calculations become unstable as soon as the reflected waves reach back the frictional interface. In any case, the pure VW friction cannot generate interface tractions that equilibrate the far field load. This law therefore lacks physical validity when the long term behavior of frictional interfaces, specially in interaction with domain boundaries, is investigated.

\section{Conclusions}	
We have discussed explicit dynamic finite element modeling of frictional interfaces governed by rate and state friction. The validity of the developed framework was assessed by comparing the numerical results with Boundary Integral Method simulations (BIM) in early stages of rupture propagation, that is before arrival of the boundary-reflected waves. We have shown that the required time step to maintain an accurate steady state sliding of the frictional interface simulated with finite element method is much smaller than the one required by BIM. This is associated to the numerical noises traveling to the interface from the internal finite element nodes and high non-linearity of the rate and state friction. Velocity weakening-strengthening (VWS) and pure velocity weakening (VW) laws both can lead to a temporary local steady state sliding on the interface behind the rupture front. However, to reach a global steady state sliding of the solid blocks, that is resulting from multiple wave reflections from the domain boundaries, the VWS law must be used. In this regard, VW can be used for short term interpretation of laboratory data, but lacks physical ground when long term behavior of frictional interfaces and interaction with domain boundaries are considered. 
\newline\\
\noindent\textbf{ACKNOWLEDGMENTS} 
Support from the Rotschild Caesarea Foundation and from the Swiss National Science Foundation (Grant No. 162569) is gratefully acknowledged. The authors wish to thank Eran Bouchbinder for discussions on rate and state friction and radiation damping.

\bibliographystyle{unsrt}
\bibliography{citations}

\begin{thebibliography}{10}

\bibitem{andrews_rupture_1976}
D.~J. Andrews.
\newblock Rupture velocity of plane strain shear cracks.
\newblock {\em Journal of Geophysical Research}, 81(32):5679--5687, November
  1976.

\bibitem{rubino_understanding_2017}
V.~Rubino, A.~J. Rosakis, and N.~Lapusta.
\newblock Understanding dynamic friction through spontaneously evolving
  laboratory earthquakes.
\newblock {\em Nature Communications}, 8:15991, June 2017.

\bibitem{marone_laboratory-derived_1998}
C.~Marone.
\newblock Laboratory derived friction laws and their application to seismic
  faulting.
\newblock {\em Annual Review of Earth and Planetary Sciences}, 26(1):643--696,
  May 1998.

\bibitem{bizzarri_slip-weakening_2003}
A.~Bizzarri and M.~Cocco.
\newblock Slip-weakening behavior during the propagation of dynamic ruptures
  obeying rate- and state-dependent friction laws.
\newblock {\em Journal of Geophysical Research}, 108(B8), 2003.

\bibitem{scholz_earthquakes_1998}
C.~H. Scholz.
\newblock Earthquakes and friction laws.
\newblock {\em Nature}, 391(6662):37--42, January 1998.

\bibitem{dieterich_constitutive_1994}
J.~H. Dieterich.
\newblock A constitutive law for rate of earthquake production and its
  application to earthquake clustering.
\newblock {\em Journal of Geophysical Research: Solid Earth},
  99(B2):2601--2618, February 1994.

\bibitem{lapusta_nucleation_2003}
N.~Lapusta and J.~R. Rice.
\newblock Nucleation and early seismic propagation of small and large events in
  a crustal earthquake model: nucleation and early seismic propagation.
\newblock {\em Journal of Geophysical Research: Solid Earth}, 108(B4), April
  2003.

\bibitem{brener_unstable_2018}
E.~A. Brener, M.~Aldam, F.~Barras, J.~F. Molinari, and E.~Bouchbinder.
\newblock Unstable slip pulses and earthquake nucleation as a nonequilibrium
  first order phase transition.
\newblock {\em Physical Review Letters}, 121(23), December 2018.

\bibitem{brener_dynamic_2016}
E.~A. Brener, M.~Weikamp, R.~Spatschek, Y.~Bar-Sinai, and E.~Bouchbinder.
\newblock Dynamic instabilities of frictional sliding at a bimaterial
  interface.
\newblock {\em Journal of the Mechanics and Physics of Solids}, 89:149--173,
  April 2016.

\bibitem{aldam_critical_2017}
M.~Aldam, M.~Weikamp, R.~Spatschek, E.~A. Brener, and E.~Bouchbinder.
\newblock Critical nucleation length for accelerating frictional slip.
\newblock {\em Geophysical Research Letters}, 44(22):11,390--11,398, November
  2017.

\bibitem{viesca_self-similar_2016}
R.~C. Viesca.
\newblock Self-similar slip instability on interfaces with rate- and
  state-dependent friction.
\newblock {\em Proceedings of the Royal Society A: Mathematical, Physical and
  Engineering Science}, 472(2192):20160254, August 2016.

\bibitem{lu_pulse-like_2010}
X.~Lu, N.~Lapusta, and A.~J. Rosakis.
\newblock Pulse-like and crack-like dynamic shear ruptures on frictional
  interfaces: experimental evidence, numerical modeling, and implications.
\newblock {\em International Journal of Fracture}, 163(1-2):27--39, May 2010.

\bibitem{lu_rupture_2010}
X.~Lu, A.~J. Rosakis, and N.~Lapusta.
\newblock Rupture modes in laboratory earthquakes: {Effect} of fault prestress
  and nucleation conditions.
\newblock {\em Journal of Geophysical Research}, 115(B12), December 2010.

\bibitem{gabriel_transition_2012}
A.~A. Gabriel, J.~P. Ampuero, L.~A. Dalguer, and P.~M. Mai.
\newblock The transition of dynamic rupture styles in elastic media under
  velocity-weakening friction.
\newblock {\em Journal of Geophysical Research: Solid Earth}, 117(B9),
  September 2012.

\bibitem{zheng-rice-1998}
G.~Zheng and J.~R. Rice.
\newblock Conditions under which velocity-weakening friction allows a
  self-healing versus a cracklike mode of rupture.
\newblock {\em Bulletin of the Seismological Society of America}, 88(6), 1998.

\bibitem{anderson_fracture_mechanics}
T.~L. Anderson.
\newblock {\em Fracture Mechanics: Fundamentals and Applications}.
\newblock CRC Press, third edition, 2005.

\bibitem{coker_frictional_2005}
D.~Coker, G.~Lykotrafitis, A.~Needleman, and A.~J. Rosakis.
\newblock Frictional sliding modes along an interface between identical elastic
  plates subject to shear impact loading.
\newblock {\em Journal of the Mechanics and Physics of Solids}, 53(4):884--922,
  April 2005.

\bibitem{shi_properties_2008}
Z.~Shi, Y.~Benzion, and A.~Needleman.
\newblock Properties of dynamic rupture and energy partition in a solid with a
  frictional interface.
\newblock {\em Journal of the Mechanics and Physics of Solids}, 56(1):5--24,
  January 2008.

\bibitem{lu_analysis_2009}
X.~Lu, N.~Lapusta, and A.~J. Rosakis.
\newblock Analysis of supershear transition regimes in rupture experiments: the
  effect of nucleation conditions and friction parameters.
\newblock {\em Geophysical Journal International}, 177(2):717--732, May 2009.

\bibitem{kaneko_spectral_2008}
Y.~Kaneko, N.~Lapusta, and J.~P. Ampuero.
\newblock Spectral element modeling of spontaneous earthquake rupture on rate
  and state faults: {Effect} of velocity-strengthening friction at shallow
  depths.
\newblock {\em Journal of Geophysical Research}, 113(B9), September 2008.

\bibitem{ampuero_cracks_2008}
J.~P. Ampuero and Y.~Ben-Zion.
\newblock Cracks, pulses and macroscopic asymmetry of dynamic rupture on a
  bimaterial interface with velocity-weakening friction.
\newblock {\em Geophysical Journal International}, 173(2):674--692, May 2008.

\bibitem{dunham_earthquake_2011}
E.~M. Dunham, D.~Belanger, L.~Cong, and J.~E. Kozdon.
\newblock Earthquake {Ruptures} with {Strongly} {Rate}-{Weakening} {Friction}
  and {Off}-{Fault} {Plasticity}, {Part} 1: {Planar} {Faults}.
\newblock {\em Bulletin of the Seismological Society of America},
  101(5):2296--2307, October 2011.

\bibitem{erickson_finite_2017}
B.~A. Erickson, E.~M. Dunham, and A.~Khosravifar.
\newblock A finite difference method for off-fault plasticity throughout the
  earthquake cycle.
\newblock {\em Journal of the Mechanics and Physics of Solids}, 109:50--77,
  December 2017.

\bibitem{allison_earthquake_2018}
K.~L. Allison and E.~M. Dunham.
\newblock Earthquake cycle simulations with rate-and-state friction and
  power-law viscoelasticity.
\newblock {\em Tectonophysics}, 733:232--256, May 2018.

\bibitem{Barras_partI}
F.~Barras, M.~Aldam, M.~Roch, E.~A. Brener, E.~Bouchbinder, and J.~F. Molinari.
\newblock The emergence of crack-like behavior of frictional rupture: The
  origin of stress drops.
\newblock {\em arxiv}, 2019.

\bibitem{Barras_partII}
F.~Barras, M.~Aldam, M.~Roch, E.~A. Brener, E.~Bouchbinder, and J.~F. Molinari.
\newblock The emergence of crack-like behavior of frictional rupture: Edge
  singularity and energy balance.
\newblock {\em To be submitted}, 2019.

\bibitem{geubelle_spectral_1995}
P.~Geubelle and J.~Rice.
\newblock A spectral method for three-dimensional elastodynamic fracture
  problems.
\newblock {\em Journal of the Mechanics and Physics of Solids},
  43(11):1791--1824, 1995.

\bibitem{barras_supershear_2018}
F.~Barras, R.~Carpaij, P.~H. Geubelle, and J.~F. Molinari.
\newblock Supershear bursts in the propagation of a tensile crack in linear
  elastic material.
\newblock {\em Physical Review E}, 98(6), December 2018.

\bibitem{barras_interplay_2017}
F.~Barras, P.~H. Geubelle, and J.~F. Molinari.
\newblock Interplay between process zone and material heterogeneities for
  dynamic cracks.
\newblock {\em Physical Review Letters}, 119(14), October 2017.

\bibitem{belytschko_nonlinear_2014}
T.~Belytschko, W.~K. Liu, B.~Moran, and K.~I. Elkhodary.
\newblock {\em Nonlinear finite elements for continua and structures}.
\newblock Wiley, Chichester, West Sussex, United Kingdon, second edition, 2014.

\bibitem{kammer_length_2016}
D.~S. Kammer, D.~Pino~Muñoz, and J.~F. Molinari.
\newblock Length scale of interface heterogeneities selects propagation
  mechanism of frictional slip fronts.
\newblock {\em Journal of the Mechanics and Physics of Solids}, 88:23--34,
  March 2016.

\bibitem{kammer_existence_2014}
D.~S. Kammer, V.~A. Yastrebov, G.~Anciaux, and J.~F. Molinari.
\newblock The existence of a critical length scale in regularised friction.
\newblock {\em Journal of the Mechanics and Physics of Solids}, 63:40--50,
  February 2014.

\bibitem{baillet_finite_2005}
L.~Baillet, V.~Linck, S.~D’Errico, B.~Laulagnet, and Y.~Berthier.
\newblock Finite {Element} {Simulation} of {Dynamic} {Instabilities} in
  {Frictional} {Sliding} {Contact}.
\newblock {\em Journal of Tribology}, 127(3):652, 2005.

\bibitem{baillet_instability_2013}
D.~Tonazzi, F.~Massi, A.~Culla, L.~Baillet, A.~Fregolent, and Y.~Berthier.
\newblock Instability scenarios between elastic media under frictional contact.
\newblock {\em Mechanical Systems and Signal Processing}, 40(2):754--766,
  November 2013.

\bibitem{baillet_instability_2007}
A.~Meziane, S.~D’Errico, L.~Baillet, and B.~Laulagnet.
\newblock Instabilities generated by friction in a pad–disc system during the
  braking process.
\newblock {\em Tribology International}, 40(7):1127--1136, July 2007.

\bibitem{laursen_constitutive_1997}
T.~A. Laursen and V.~G. Oancea.
\newblock On the constitutive modeling and finite element computation of
  rate-dependent frictional sliding in large deformations.
\newblock {\em Computer Methods in Applied Mechanics and Engineering},
  143(3-4):197--227, April 1997.

\bibitem{oancea_dynamics_1996}
V.~G. Oancea and T.~A. Laursen.
\newblock Dynamics of a state variable frictional law in finite element
  analysis.
\newblock {\em Finite Elements in Analysis and Design}, 22(1):25--40, May 1996.

\bibitem{tal_dynamic_2018}
Y.~Tal and B.~H. Hager.
\newblock Dynamic mortar finite element method for modeling of shear rupture on
  frictional rough surfaces.
\newblock {\em Computational Mechanics}, 61(6):699--716, June 2018.

\bibitem{dieterich_modeling_1979}
J.~H. Dieterich.
\newblock Modeling of rock friction: 1. {Experimental} results and constitutive
  equations.
\newblock {\em Journal of Geophysical Research}, 84(B5):2161, 1979.

\bibitem{ruina_slip_1983}
A.~Ruina.
\newblock Slip instability and state variable friction laws.
\newblock {\em Journal of Geophysical Research: Solid Earth},
  88(B12):10359--10370, December 1983.

\bibitem{Dieterich_direct_observ_1994}
J.~H. Dieterich and B.~D. Kilgore.
\newblock Direct observation of frictional contacts: New insights for
  state-dependent properties.
\newblock {\em Pure and Applied Geophysics}, 143(1-3):283--302, 1994.

\bibitem{Dieterich_1986}
J.~H. Dieterich.
\newblock A model for the nucleation of earthquake slip.
\newblock {\em Das et al Geophysical Monograph}, 37, 1986.

\bibitem{Okubo_Dieterich_1986}
P.~G. Okubo and J.~H. Dieterich.
\newblock State variable fault constitutive relations for dynamic slip.
\newblock {\em Das et al Geophysical Monograph 37: Earthquake Source
  Mechanics}, 1986.

\bibitem{shibazaki_physical_2003}
B.~Shibazaki and Y.~Iio.
\newblock On the physical mechanism of silent slip events along the deeper part
  of the seismogenic zone.
\newblock {\em Geophysical Research Letters}, 30(9), 2003.

\bibitem{sinai_velocity-strengthening_2014}
Y.~Bar-Sinai, R.~Spatschek, E.~A. Brener, and E.~Bouchbinder.
\newblock On the velocity-strengthening behavior of dry friction.
\newblock {\em Journal of Geophysical Research: Solid Earth},
  119(3):1738--1748, March 2014.

\bibitem{nagata_revised_2012}
K.~Nagata, M.~Nakatani, and S.~Yoshida.
\newblock A revised rate- and state-dependent friction law obtained by
  constraining constitutive and evolution laws separately with laboratory data.
\newblock {\em Journal of Geophysical Research: Solid Earth}, 117(B2):n/a--n/a,
  February 2012.

\bibitem{Barras_Thesis}
F.~Barras.
\newblock When dynamic cracks meet disorder: A journey along the fracture
  process zone (phd thesis).
\newblock {\em Ecole Polytechnique Fédérale de Lausanne (EPFL)}, (Thesis No.
  8956), 2018.

\bibitem{curnier_computational_1994}
A.~Curnier.
\newblock {\em Computational {Methods} in {Solid} {Mechanics}}, volume~29 of
  {\em Solid {Mechanics} and {Its} {Applications}}.
\newblock Springer Netherlands, Dordrecht, 1994.

\end{thebibliography}

\begin{appendices}
\numberwithin{equation}{section}

\section{Time discretization procedure}\label{time-disc-proc}
Various time discretization methods are available in the literature to solve Equation \ref{motion-3}. In this paper, we employ the central difference method, which is an explicit time integration scheme \cite{curnier_computational_1994}. All dynamic quantities at the time $t_{n+1} = t_n + \Delta t$ are calculated in terms of the known quantities at previous step $t_n$. Having displacement ${\mbf u}_n$, velocity $\dot{\mbf u}_n$, and acceleration $\ddot{\mbf u}_n$ vectors calculated at time $t_n$, one can obtain displacement and predicted velocity vectors at $t_{n+1}$ as
\begin{equation} \label{u_n+1}
{\mbf u}_{n+1} = {\mbf u}_n + \Delta t~\dot{\mbf u}_n + \frac{{\Delta t}^2}{2} \ddot{\mbf u}_n
\end{equation}
\begin{equation} \label{vp_n+1}
\dot{\mbf u}^{\text{p}}_{n+1} = \dot{\mbf u}_n + \Delta t~\ddot{\mbf u}_n
\end{equation}
where ``p" superscript denotes predicted value at $t_{n+1}$, which will be corrected upon calculation of acceleration increment over $t_{n+1} - t_n$. Next, the acceleration vector at $t_{n+1}$ is 
\begin{equation} \label{a_n+1}
\ddot{\mbf u}_{n+1} = \mbf{M}^{-1} \bigg(\mbf{f}^{\text{ext}}_{n+1} - \mbf{f}^{\text{int}}_{n+1}  +  \mbf{f}^{\text{cont}}_{n+1} + \mbf{f}^{\text{fric}}_{n+1} \bigg).
\end{equation}

Considering that ${\mbf u}_{n+1}$, calculated by Equation \ref{u_n+1}, is employed for the calculation of internal force vector $\mbf{f}^{\text{int}}_{n+1}$, one can rewrite Equation \ref{a_n+1} as
\begin{equation} \label{a_n+1_free}
\ddot{\mbf u}_{n+1} = \ddot{\mbf u}^{\text{free}}_{n+1} + \mbf{M}^{-1} \bigg(\mbf{f}^{\text{cont}}_{n+1} + \mbf{f}^{\text{fric}}_{n+1} \bigg)
\end{equation}
in which $\ddot{\mbf u}^{\text{free}}_{n+1} = \mbf{M}^{-1} \bigg(\mbf{f}^{\text{ext}}_{n+1} - \mbf{f}^{\text{int}}_{n+1} \bigg)$ is the acceleration vector at $t_{n+1}$ without considering interfacial contact and friction. Derivation of contact and friction force vectors, $\mbf{f}^{\text{cont}}_{n+1}$ and $\mbf{f}^{\text{fric}}_{n+1}$, are presented in the next two sections. Having $\ddot{\mbf u}_{n+1}$ calculated, the acceleration increment is computed as $\delta \ddot{\mbf u}_{n+1} = \ddot{\mbf u}_{n+1} - \ddot{\mbf u}_{n}$, and the predicted velocity vector, Equation \ref{vp_n+1}, is corrected as explained in \cite{curnier_computational_1994}
\begin{equation} \label{v_n+1}
\dot{\mbf u}_{n+1} = \dot{\mbf u}^{\text{p}}_{n+1} + \frac{\Delta t}{2}~\delta \ddot{\mbf u}_{n+1}.
\end{equation}

\section{Derivation of the normal traction vector $\mbf{f}^{\text{cont}}$ on the interface}\label{contact-force-derivation}
The main challenge of the integration procedure is the calculation of contact and friction force vectors on the interface in Equation \ref{a_n+1}. In this paper, a node to node contact algorithm is employed, which we detail below. It should be noted that the node to node contact algorithm is only valid for small displacements, which is the case for the numerical examples presented in this paper. The normal contact force vector on the interface is calculated through enforcing normal gap value equal or greater than zero for all interfacial node pairs. Considering a generic node pair $i$ on the interface, the interaction between the two nodes is illustrated in Figure \ref{FE-contact-setup} in an enlarged view. The two nodes belong to the positive and negative contact surfaces, $\Gamma_c^{+}$ and $\Gamma_c^{-}$, and the normal gap constraint between them is enforced through
\begin{equation}\label{A.1}
\llbracket x_{n+2} \rrbracket^N \geq 0
\end{equation}
where $\llbracket x \rrbracket = x^+ - x^-$ represents jump of variable $x$, which is the current nodal position value at time $t_{n+2}$. The superscript $N$ represents the component that is normal to the interface. The index $i$, corresponding to the interfacial node pair $i$, is dropped from the derivations to simplify notations. Replacing $\llbracket x_{n+2} \rrbracket = \llbracket x_{n+1} \rrbracket +  \llbracket u_{n+2} \rrbracket$ along with Equation \ref{u_n+1} in Equation \ref{A.1} gives
\begin{equation}\label{A.2}
\llbracket x_{n+1} \rrbracket^N  + \llbracket {u}_{n+1} \rrbracket^N  + \Delta t~ \llbracket \dot{u}_{n+1} \rrbracket^N  + \frac{{\Delta t}^2}{2} \llbracket \ddot{u}_{n+1} \rrbracket ^N \geq 0.
\end{equation}

Replacing Equation \ref{v_n+1} in Equation \ref{A.2} yields
\begin{equation}\label{A.3} 
\llbracket x_{n+1} \rrbracket^N  + \llbracket {u}_{n+1} \rrbracket^N  + \Delta t~ \llbracket \dot{u}^{\text{p}}_{n+1} \rrbracket^N + \frac{{\Delta t}^2}{2}~\llbracket \delta \ddot{u}_{n+1} \rrbracket^N  + \frac{{\Delta t}^2}{2} \llbracket \ddot{u}_{n+1} \rrbracket ^N \geq 0.
\end{equation}

Using $\llbracket \delta \ddot{u}_{n+1} \rrbracket^N = \llbracket \ddot{u}_{n+1} \rrbracket^N - \llbracket \ddot{u}_{n} \rrbracket^N$, we obtain
\begin{equation}\label{A.4} 
\llbracket x_{n+1} \rrbracket^N  + \llbracket {u}_{n+1} \rrbracket^N  + \Delta t~ \llbracket \dot{u}^{\text{p}}_{n+1} \rrbracket^N - \frac{{\Delta t}^2}{2}~\llbracket \ddot{u}_{n} \rrbracket^N  +{\Delta t}^2 \llbracket \ddot{u}_{n+1} \rrbracket^N \geq 0,
\end{equation}

which considering Equation \ref{a_n+1_free} gives
\begin{equation}\label{A.5} 
\llbracket x_{n+1} \rrbracket^N  + \llbracket {u}_{n+1} \rrbracket^N  + \Delta t~ \llbracket \dot{u}^{\text{p}}_{n+1} \rrbracket^N - \frac{{\Delta t}^2}{2}~\llbracket \ddot{u}_{n} \rrbracket^N  + {\Delta t}^2 \llbracket \ddot{u}^{\text{free}}_{n+1} \rrbracket^{N} + {\Delta t}^2 {M}^{-1} \llbracket B R^N_{n+1} \rrbracket \geq 0
\end{equation}
where $B R^N_{n+1} = f^{\text{cont}}_{n+1}$, in which $R^N_{n+1}$ is the interface traction in normal direction, and $B$ is the area associated to the contact node on the interface, see Figure \ref{FE-contact-setup}. Since the contact forces on the interface applied on the two surfaces are equal and in opposite directions: $R^{N+}_{n+1} = -R^{N-}_{n+1} = -R^{N}_{n+1}$, one can write $\llbracket B R^N_{n+1} \rrbracket = -2B R^N_{n+1}$. Introducing an aggreagate variable $Z$ \cite{kaneko_spectral_2008}: 
\begin{equation}\label{A.6} 
Z^{-1} = \frac{\Delta t}{2}(M^{-1}_{+}B_{+} + M^{-1}_{-}B_{-}),
\end{equation}
equation \ref{A.5} yields to
\begin{equation}\label{A.7} 
\llbracket x_{n+1} \rrbracket^N  + \llbracket {u}_{n+1} \rrbracket^N  + \Delta t~ \llbracket \dot{u}^{\text{p}}_{n+1} \rrbracket^N - \frac{{\Delta t}^2}{2}~\llbracket \ddot{u}_{n} \rrbracket^N  + {\Delta t}^2 \llbracket \ddot{u}^{\text{free}}_{n+1} \rrbracket^{N} - 2{\Delta t}~Z^{-1}R^{N}_{n+1} \geq 0.
\end{equation}

Finally, we define
\begin{equation}\label{A.8} 
I_{n+1}^{\text{free}} = \llbracket x_{n+1} \rrbracket^N  + \llbracket {u}_{n+1} \rrbracket^N  + \Delta t~ \llbracket \dot{u}^{\text{p}}_{n+1} \rrbracket^N - 0.5{\Delta t}^2~\llbracket \ddot{u}_{n} \rrbracket^N  + {\Delta t}^2~\llbracket \ddot{u}^{\text{free}}_{n+1} \rrbracket^{N}
\end{equation}
as the interpenetration between the two surfaces, and we can evaluate the normal contact force between the two corresponding nodes as follows
\begin{equation}\label{A.9} 
f^{\text{cont}}_{n+1} = B R^N_{n+1}; \hspace{0.5 cm}
R^{N}_{n+1} = 
\begin{cases}
    \frac{1}{2\Delta t} Z I_{n+1}^{\text{free}} \hspace{1cm} \text{if~~} I_{n+1}^{\text{free}} \leq 0\\
    0  \hspace{2 cm} \text{if~~} I_{n+1}^{\text{free}} > 0.
\end{cases}
\end{equation}

\section{Derivation of the tangential traction vector $\mbf{f}^{\text{fric}}$ on the interface}\label{friction-force-derivation}
To calculate the tangential traction on the interface, the relative velocity between every pair of nodes on the interface at mid time step is equated to zero as   
\begin{equation}\label{B.1}
\llbracket \dot{u}_{n+3/2} \rrbracket^T = 0
\end{equation}
in which the superscript $T$ represents the component that is tangential to the interface. Considering $\dot{u}_{n+3/2} = \dot{u}_{n+1} + 0.5\Delta t~\ddot{u}_{n+1}$ along with Equation \ref{v_n+1} gives
\begin{equation}\label{B.2}
\llbracket \dot{u}^{\text{p}}_{n+1} \rrbracket^T + \frac{\Delta t}{2}~\llbracket \delta \ddot{u}_{n+1} \rrbracket^T + \frac{\Delta t}{2}~\llbracket \ddot{u}_{n+1} \rrbracket^T = 0,
\end{equation}
which can be simplified to
\begin{equation}\label{B.3}
\llbracket \dot{u}^{\text{p}}_{n+1} \rrbracket^T - \frac{\Delta t}{2}~\llbracket \ddot{u}_{n} \rrbracket^T + \Delta t~\llbracket \ddot{u}_{n+1} \rrbracket^T = 0.
\end{equation}

Using Equation \ref{a_n+1_free} and defining $B R^T_{n+1} = f^{\text{fric}}_{n+1}$
\begin{equation}\label{B.4}
\llbracket \dot{u}^{\text{p}}_{n+1} \rrbracket^T - \frac{\Delta t}{2}~\llbracket \ddot{u}_{n} \rrbracket^T + {\Delta t} \llbracket \ddot{u}^{\text{free}}_{n+1} \rrbracket^{T} + {\Delta t} {M}^{-1} \llbracket B R^T_{n+1} \rrbracket = 0.
\end{equation}

Similar to the derivation of the normal traction, $\llbracket B R^T_{n+1} \rrbracket = -2B R^T_{n+1}$, so Equation \ref{B.4} can be written as 
\begin{equation}\label{B.5}
\llbracket \dot{u}^{\text{p}}_{n+1} \rrbracket^T - \frac{\Delta t}{2}~\llbracket \ddot{u}_{n} \rrbracket^T + {\Delta t} \llbracket \ddot{u}^{\text{free}}_{n+1} \rrbracket^{T} -2 Z^{-1} R^T_{n+1} = 0.
\end{equation}

Defining 
\begin{equation}\label{B.6}
T_{n+1}^{\text{free}} = \llbracket \dot{u}^{\text{p}}_{n+1} \rrbracket^T - 0.5{\Delta t}~\llbracket \ddot{u}_{n} \rrbracket^T + {\Delta t}~\llbracket \ddot{u}^{\text{free}}_{n+1} \rrbracket^{T}
\end{equation}
one obtains the tangential traction between the pair of nodes as
\begin{equation}\label{B.7}
R^T_{n+1} = \frac{1}{2} Z T_{n+1}^{\text{free}}
\end{equation}

The tangential traction calculated through Equation \ref{B.7} is the necessary traction to hinder relative velocity between the considered pair of nodes. Next, this traction must be compared to the frictional strength calculated through rate and state friction. The coefficient of friction $f(v,\phi)$ can be calculated using Equations \ref{RS-VW} and \ref{state-diff-eq}, in which $v=\llbracket \dot{u}^{\text{p}}_{n+1} \rrbracket^T$ is the relative slip velocity computed at the corresponding node pair. Finally, the frictional force can be calculated as
 
\begin{equation}\label{B.8} 
f^{\text{fric}}_{n+1} = B R^T_{n+1}; \hspace{0.5 cm}
R^{T}_{n+1} = 
\begin{cases}
     R^{T}_{n+1}                \hspace{1.9cm}   \text{if	~~} R^{T}_{n+1} < f(v,\phi) R^{N}_{n+1}\\
     f(v,\phi) R^{N}_{n+1}  \hspace{1cm}      \text{if	~~} R^{T}_{n+1} \geq f(v,\phi) R^{N}_{n+1}
\end{cases}
\end{equation}

\section{Numerical procedure steps}\label{num-steps}
All necessary equations for the explicit dynamic finite element modeling of frictional interface between two elastic bodies having been presented, we now detail the sequence of calculations carried at a given time step $\Delta t = t_{n+1}-t_{n}$. Given that displacement ${\mbf u}_n$, velocity $\dot{\mbf u}_n$, and acceleration $\ddot{\mbf u}_n$ vectors are known at $t_{n}$, we aim to calculate these quantities at $t_{n+1}$:
\begin{enumerate}[label={(\arabic*)}]
\item ${\mbf u}_{n+1}$ and $\dot{\mbf u}^{\text{p}}_{n+1}$ are calculated using Equations \ref{u_n+1} and \ref{vp_n+1}.

\item Having external forces at $t_{n+1}$, $\mbf{f}^{\text{ext}}_{n+1}$, and calculating the internal forces, $\mbf{f}^{\text{int}}_{n+1}$ by using ${\mbf u}_{n+1}$ along with the material constitutive equation, one can then calculate $\ddot{\mbf u}^{\text{free}}_{n+1} = \mbf{M}^{-1} \bigg(\mbf{f}^{\text{ext}}_{n+1} - \mbf{f}^{\text{int}}_{n+1} \bigg)$.

\item A loop over all the node pairs at the interface is carried out to calculate $f^{\text{cont}}_{n+1}$ and $f^{\text{fric}}_{n+1}$:

\begin{itemize}
\item $I_{n+1}^{\text{free}}$ is calculated using the jump of calculated quantities in Equation \ref{A.8}. $R^{N}_{n+1}$ and subsequently $f^{\text{cont}}_{n+1}$ are calculated through Equation \ref{A.9}.

\item The jump of calculated quantities are used in Equation \ref{B.6} to obtain $T_{n+1}^{\text{free}}$, which is then used in Equation \ref{B.7} to obtain $R^T_{n+1}$. This value should be compared to the frictional strength calculated through rate and state friction. 
\begin{itemize}
\item Having $\phi_n$ and discretizing Equation \ref{state-diff-eq} in time, the state variable at $t_{n+1}$ can be obtained through $\phi_{n+1} = \phi_{n} + (1 - v_{n+1} \phi_n/D)\Delta t$ where $v_{n+1}=\llbracket \dot{u}^{\text{p}}_{n+1} \rrbracket^T$. Next, the coefficient of friction $f(v_{n+1},\phi_{n+1})$ is calculated through the considered friction law, Equation \ref{RS-VW} or \ref{RS-VWS}.
\end{itemize}

$R^T_{n+1}$ is then compared to $f(v_{n+1},\phi_{n+1}) R^N_{n+1}$ as explained in Equation \ref{B.8} to obtain $f^{\text{fric}}_{n+1}$.
\end{itemize}

\item Shaping the contact and friction forces calculated for each node pair in vector format, $\mbf{f}^{\text{cont}}_{n+1}$ and $\mbf{f}^{\text{fric}}_{n+1}$, Equation \ref{a_n+1_free} is then used to calculate $\ddot{\mbf u}_{n+1}$.

\item Finally, the velocity values are corrected through Equation \ref{v_n+1}, and $\dot{\mbf u}_{n+1}$ is obtained.

\end{enumerate}


\end{appendices}

\end{document}